\documentclass[noacm,sigconf,balance=false]{acmart}
\renewcommand\footnotetextcopyrightpermission[1]{}

\usepackage{subfig}
\usepackage{enumitem}
\usepackage{algorithm,algpseudocode}
\usepackage{tabularx}
\usepackage{amsfonts,amsmath,amsthm,stmaryrd}
\usepackage{tcolorbox}
\usepackage{url,xurl,hyperref}

\newtheorem{Thm}{Theorem}[section]
\newtheorem{Lem}[Thm]{Lemma}

\newtheorem{Cor}[Thm]{Corollary}

\newtheorem{Def}[Thm]{Definition}

\def\R{\mathbb{R}}

\def\N{\mathbb{N}}

\def\F{\mathbb{F}}

\usepackage{bbm}
\def\1{\mathbbm{1}}
\renewcommand{\epsilon}{\varepsilon}

\newcommand{\prob}[2][]{\Pr\ifstrempty{#1}{}{_{#1}}\left[\begin{array}{@{}c@{}}#2\end{array}\right]}

\newcommand{\case}[1]{\left\{\begin{array}{@{}ll@{}}#1\end{array}\right.}

\usepackage[
	lambda,
	operators,
	advantage,
	sets,
	adversary,
	landau,
	probability,
	notions,	
	logic,
	ff,
	mm,
	primitives,
	events,
	complexity,
	oracles,
	asymptotics,
	keys]{cryptocode}
\usepackage{adjustbox}

\renewcommand{\sf}[1]{\ifmmode\mathsf{#1}\else\textsf{#1}\fi}
\newcommand{\adbf}[1]{\ifmmode\mathbf{#1}\else\textbf{#1}\fi}

\def\1{\mathbbm{1}}

\def\ser{\ensuremath{\sf{Ser}}}
\def\cli{\ensuremath{\sf{Cli}}}

\usepackage{multirow}

\usepackage{booktabs}

\usepackage{pifont}

\newcommand{\autofit}[1]{\maxsizebox{\linewidth}{!}{\begin{varwidth}{10\linewidth}#1\end{varwidth}}}

\allowdisplaybreaks

\usepackage{etoolbox}



\begin{document}

\title{Robust, Lightweight Locally Differentially Private Histogram Estimation and the Role of Dropout Attacks}

\author{Haochen Sun}
\authornote{Haochen Sun and Jacob Imola contributed equally to this work.}
\affiliation{%
  \institution{University of Waterloo}
  \city{Waterloo}
  \country{Canada}}
\email{haochen.sun@uwaterloo.ca}

\author{Jacob Imola}
\authornotemark[1]
\affiliation{%
  \institution{University of Waterloo}
  \city{Waterloo}
  \country{Canada}}
\email{jimola@uwaterloo.ca}

\author{Xi He}
\affiliation{%
  \institution{University of Waterloo}
  \institution{Vector Institute}
  \city{Waterloo}
  \country{Canada}}
\email{xi.he@uwaterloo.ca}

\renewcommand{\shortauthors}{Sun, Imola, and He}

\begin{abstract}
    Local differential privacy (LDP) is vulnerable to output manipulation by malicious clients.
    Zero-knowledge proofs can enforce correct report computation, but clients may still selectively drop out after inspecting their outputs.
    Prior work combines zero-knowledge proofs with homomorphic encryption to hide reports from clients and prevent selective dropout, but incurs substantial computational cost.
    We instead develop a three-message lightweight verification framework for robust histogram estimation that uses proofs of plaintext evaluation to restrict output manipulation to selective dropout.
    Across the evaluated settings for PI-RAPPOR and Hadamard Response, mean per-client online latency remains below 500 milliseconds.
    We further establish tight error bounds showing that, for RAPPOR, PI-RAPPOR, and Hadamard Response, restricting malicious clients to selective dropout reduces attack-dependent error by a factor proportional to the square root of the domain size compared with output manipulation.
    Additionally, when client protocols cannot be modified, we propose server-side consistency checks for RAPPOR and Hadamard Response.
    These checks improve robustness without client-side cryptography and can be applied post-hoc to collected reports, offering an alternative under different deployment constraints.
    Experiments demonstrate that both verification and consistency checks substantially reduce attack error for certain mechanisms, with consistency checks yielding errors intermediate between those obtained with verification and those under unrestricted output manipulation.
\end{abstract}

\maketitle
\pagestyle{plain}

\section{Introduction}
\label{sec:introduction}

Local differential privacy (LDP) enables clients to report sensitive data to an untrusted server by randomizing their inputs locally~\cite{DworkMNS06,DworkR14,EvfimievskiGS03}.
We study histogram estimation, where each client holds a categorical value and the server estimates category counts from randomized reports, a basic task in locally private data collection~\cite{WangBLJ17}.
Malicious clients can bypass the prescribed randomizer and submit arbitrary reports, substantially distorting the estimated histogram~\cite{CheuSU21,CaoJG21}.
Verifiable LDP protocols address this threat by requiring clients to prove that their reports were generated according to the prescribed computation.

Correct report generation, however, does not ensure unbiased submission.
Even when a protocol enforces correct randomization, a malicious client that can inspect its report may submit it only when the result is favorable.
Otherwise, it can withhold the report or send an invalid proof that the server rejects.
An adversary controlling multiple clients can coordinate these decisions to bias the aggregate.
All accepted reports may therefore be valid, while their distribution is altered by \emph{selective dropout}.
The proof certifies how a submitted report was computed, but not why the client chose to submit it.

The V$\epsilon$rity protocol~\cite{BellClarkGLRC25} prevents selection based on the realized report by hiding it from the client.
It supports client-chosen inputs in a single server--client--server round trip.
The server supplies encrypted randomness, and the client evaluates the mechanism homomorphically, returning an encrypted report with zero-knowledge proofs of the required cryptographic operations.
The client cannot inspect the report before deciding whether to submit it.
The computational cost is substantial, however: in the reported binary randomized response configuration, client and server computation each exceed 20 minutes.
This raises our central question:
\emph{Is eliminating selective dropout necessary to obtain meaningful robustness, or can a lightweight protocol retain a substantial benefit while allowing it?}

The answer depends on the underlying LDP mechanism.
If selective dropout and unrestricted \emph{output manipulation} have the same asymptotic effect, restricting malicious clients to selective dropout does not improve the asymptotic error rate.
Binary randomized response illustrates this difficulty, but does not settle the question for larger categorical domains.
We show that RAPPOR~\cite{ErlingssonPK14}, PI-RAPPOR~\cite{FeldmanT21}, and Hadamard Response~\cite{AcharyaSZ19} exhibit an asymptotic separation:
the attack-dependent term in the worst-case expected $\ell_1$ error under selective dropout is a factor of $\sqrt{k}$ smaller than under output manipulation, where $k$ is the number of categories.
In contrast, HST~\cite{BassilyNST20,CheuSU21} has the same asymptotic error rate under both attacks.
Thus, verification can improve robustness for (PI-)RAPPOR and Hadamard Response even when selective dropout remains possible.

This separation motivates a lightweight verification framework that prevents output manipulation while allowing selective dropout.
The client first commits to its input and randomness share.
The server then supplies an independent randomness share, after which the client returns its report and a zero-knowledge proof of correct evaluation.
The commitment fixes the client's choices before the server's share is revealed, preventing the client from repeatedly generating reports and selecting a favorable one.
The resulting three-message, 1.5-round-trip protocol requires proofs of plaintext computation rather than homomorphic operations.
Clients can still withhold their fixed reports, but our analysis quantifies the resulting error.
We instantiate the framework for PI-RAPPOR and Hadamard Response using compact verification circuits.

When adding verification to the client is impractical, we pursue a complementary route: server-side consistency checks that reduce the effects of output manipulation without changing the client protocol.
For RAPPOR, the check detects unusually large pairwise correlations between reports.
For Hadamard Response, it clips unusually large counts in the report histogram.
These checks can be applied post-hoc to collected reports.
They require no client-side cryptography, though their error guarantees are weaker than those under selective dropout with verification.

Table~\ref{tab:comparison} summarizes the error guarantees and aggregation costs.
HST provides a useful contrast: its error rate under output manipulation matches the selective-dropout rate of the verified mechanisms asymptotically, while its explicit aggregation procedure takes $O\left(nk\right)$ time.
Each HST client sends one bit and uses $k$ public random bits; explicitly distributing these vectors incurs $O\left(k\right)$ downlink communication per client.
The underlying PI-RAPPOR and Hadamard Response mechanisms have logarithmic-size reports and admit $O\left(n+k\log k\right)$ aggregation.
Our verification protocols introduce additional cryptographic messages, whose full communication costs and contribution to modeled latency are evaluated in Section~\ref{sec:exp}.

We summarize our contributions as follows:

\begin{itemize}[leftmargin=*]
    \item \textbf{A lightweight verification framework and instantiations.}
    We formalize a resampling vulnerability in one-round-trip protocols with client-visible reports: under suitable conditions, clients can repeatedly generate valid reports and submit a favorable one \textbf{(Section~\ref{sec:ideal-impossibility})}.
    We give a three-message framework that fixes the client's input and randomness before the server reveals its share, preventing resampling while allowing selective dropout.
    We instantiate the framework for PI-RAPPOR and Hadamard Response using proofs of plaintext evaluation \textbf{(Section~\ref{sec:protocol})}.

    \item \textbf{A mechanism-dependent separation between attacks.}
    We establish tight error bounds under selective dropout for RAPPOR, PI-RAPPOR, Hadamard Response, and HST.
    For the first three mechanisms, the attack-dependent term in the worst-case expected $\ell_1$ error is smaller by a factor of $\sqrt{k}$ than under output manipulation.
    HST exhibits no such asymptotic separation.
    These results identify when the restriction enforced by verification improves robustness \textbf{(Section~\ref{sec:dropout-analysis})}.

    \item \textbf{Robustness without cryptography.}
    We develop post-hoc consistency checks for RAPPOR and Hadamard Response, with error bounds under output manipulation.
    The checks operate entirely at the server and require no changes to the client protocol \textbf{(Section~\ref{sec:cons})}.

    \item \textbf{Empirical robustness and efficient verification.}
    Our experiments illustrate the mechanism-dependent separation between attacks and evaluate the consistency checks.
    Our PI-RAPPOR and Hadamard Response implementations achieve mean modeled per-client online latency below $0.5$ seconds across the evaluated settings, combining computation measured on server hardware with modeled network transmission.
    Additionally, in our experiments at $\epsilon=1$ with 10\% corrupted clients, HST under output manipulation has up to 2.7 times the mean error of verified PI-RAPPOR under selective dropout and up to 2.6 times that of verified Hadamard Response under selective dropout (Section~\ref{sec:exp}).
\end{itemize}

\begin{table}[t]
    \centering
    \setlength{\tabcolsep}{3pt}
    \caption{
        Histogram estimation for $0<\epsilon\leq 1$.
        OM: output manipulation; SD: selective dropout;
        CC: consistency check; Agg.: baseline aggregation.
        HR denotes Hadamard Response.
        Error entries omit a common factor $\frac{1}{\epsilon}$
        and the honest-sampling term
        $O\left(\frac{k\sqrt{n}}{\epsilon}\right)$.
        OM and SD give tight rates; CC gives upper bounds
        under the assumptions of Section~\ref{sec:cons}.
        Aggregation entries denote $O\left(\cdot\right)$ costs,
        excluding verification and consistency checks.
        HST uses explicit public vectors.
    }
    \label{tab:comparison}
    \begin{tabular}{lcccc}
        \hline
        Protocol & OM & SD & CC & Agg. \\
        \hline
        RAPPOR
        & $mk$
        & $m\sqrt{k}$
        & $mk^{\frac{3}{4}}(\log n)^{\frac{1}{4}}$
        & $nk$ \\
        PI-RAPPOR
        & $mk$
        & $m\sqrt{k}$
        & ---
        & $n+k\log k$ \\
        HR
        & $mk$
        & $m\sqrt{k}$
        & $\sqrt{km\left(n+k\log(nk)\right)}$
        & $n+k\log k$ \\
        HST
        & $m\sqrt{k}$
        & $m\sqrt{k}$
        & ---
        & $nk$ \\
        \hline
    \end{tabular}
\end{table}

\subsection{Related Work}

\noindent\textbf{Manipulation attacks and defenses for LDP.}
Cheu, Smith, and Ullman~\cite{CheuSU21} distinguish input manipulation, in which clients randomize chosen false inputs, from attacks that manipulate reports directly.
They show that LDP can amplify manipulation and that robustness depends on the mechanism, including HST.
Cao, Jia, and Gong~\cite{CaoJG21} study input and output poisoning attacks on frequency estimation and heavy-hitter protocols, and evaluate countermeasures.
Song, Xu, and Zhu~\cite{SongXZ23} propose interactive verifiable defenses against output poisoning for generalized randomized response and optimized unary encoding.
Our work builds on these studies by establishing tight selective-dropout error bounds and developing server-side consistency checks for RAPPOR and Hadamard Response.

\noindent\textbf{Verifiable LDP.}
Kato et al.~\cite{KatoCY21} propose interactive protocols for verifiable randomized response.
Bontekoe et al.~\cite{BontekoeAT25} give verifiable local and shuffle protocols assuming authenticated raw inputs.
Most relevant to our work, V$\epsilon$rity~\cite{BellClarkGLRC25} provides verifiable binary randomized response for client-chosen inputs and hides reports using HE and proofs of HE operations in a single server--client--server round trip.
The same work analyzes dropout attacks in verifiable LDP.
For $n$ users and $m$ malicious users, V$\epsilon$rity allows client-chosen inputs but prevents selection based on the noise, giving an error bound of $\widetilde{O}\left(m+\frac{\sqrt{n}}{\epsilon}\right)$.
We instead allow clients to observe their reports and quantify the resulting selective-dropout error for categorical histogram mechanisms.
Our analysis identifies when verification improves robustness over unrestricted output manipulation despite selective dropout, motivating verification through plaintext rather than homomorphic evaluation.

\noindent\textbf{Verifiable central and distributed DP.}
Verifiable DP has also been studied in central and distributed settings~\cite{NarayanFPH15}.
Biswas and Cormode~\cite{BiswasC23} give interactive zero-knowledge proofs for differentially private counting with the binomial mechanism, combining committed private randomness with public randomness.
Confidential-DPproof~\cite{ShamsabadiTCBHP24} certifies differentially private training without revealing the training data or model.
Noisette~\cite{PangGLXSZW26} develops certifiable lookup-table evaluation for discrete DP noise sampling and extends its approach to continuous sampling with end-to-end privacy guarantees under floating-point arithmetic.
Sun and He~\cite{SunH25} formalize verifiable distributed DP in a client--server--verifier model and construct a verifiable distributed discrete Laplace mechanism, accounting for collusion and finite-precision sampling.
\section{Preliminaries}\label{sec:prelim}

\subsection{Local Differential Privacy}

Let $\mathcal{X}=\{0,\ldots,k-1\}$ be the data domain and let $\mathcal{Y}$ be the output space of a local randomizer.
We write $\mathcal{M}(x)\in\mathcal{Y}$ for the output of a fresh execution of the randomized mechanism $\mathcal{M}$ on input $x\in\mathcal{X}$.

\begin{Def}[$\epsilon$-local differential privacy (LDP)~\cite{DworkMNS06, EvfimievskiGS03}]\label{def:ldp}
    A randomized mechanism $\mathcal{M}$ with input domain $\mathcal{X}$ and output space $\mathcal{Y}$ satisfies $\epsilon$-local differential privacy if, for every pair $x,x'\in\mathcal{X}$ and every measurable set $S\subseteq\mathcal{Y}$,
    \begin{equation}
        \prob{\mathcal{M}(x)\in S}
        \leq e^\epsilon\prob{\mathcal{M}(x')\in S}.
    \end{equation}
\end{Def}

For a cryptographic protocol, the server observes not only the LDP report but also commitments, proofs, and other protocol messages.
We therefore use the computational analogue of LDP~\cite{MironovPRV09} to reason about the server's entire view.

\begin{Def}[$\epsilon$-computational local differential privacy~\cite{MironovPRV09, EvfimievskiGS03}]
    \label{def:computational-ldp}
    Let $\mathsf{View}_{\ser}(x)$ denote the server's view of an execution with an honest client holding input $x$.
    The protocol satisfies $\epsilon$-computational local differential privacy if, for every pair $x,x'\in\mathcal{X}$ and every PPT distinguisher $\mathcal{A}$, there is a negligible function $\nu$ such that
    \begin{equation}
        \begin{aligned}
            &\prob{\mathcal{A}(\mathsf{View}_{\ser}(x))=1}\\
            &\quad\leq e^\epsilon
            \prob{\mathcal{A}(\mathsf{View}_{\ser}(x'))=1}
            +\nu(\kappa).
        \end{aligned}
        \label{eq:computational-ldp}
    \end{equation}
\end{Def}

In the local model, client $i$ applies $\mathcal{M}$ to its own value $x_i$ and sends only the resulting report $y_i$ to the server.
Thus, the privacy guarantee holds before reports from different clients are aggregated.
Our statistical task is to estimate frequencies from the clients' reports.
For inputs $\mathbf{x}=(x_0,\ldots,x_{n-1})\in\mathcal{X}^n$, the true histogram is
\begin{equation}
    F(\mathbf{x})=\sum_{i=0}^{n-1} \mathbf{e}_{x_i}\in\R^k, \label{eq:true-histogram}
\end{equation}
where $\mathbf{e}_j$ denotes the $j$th standard basis vector.
Unless stated otherwise, each honest client uses fresh, independent randomness when executing $\mathcal{M}$.
Logarithms are natural unless a base is specified.

We study four representative LDP mechanisms for frequency estimation over large domains: RAPPOR~\cite{ErlingssonPK14}, PI-RAPPOR~\cite{FeldmanT21}, Hadamard Response~\cite{AcharyaSZ19}, and HST~\cite{BassilyNST20,CheuSU21}.
With RAPPOR, each client independently randomizes a one-hot vector and sends the result; RAPPOR was deployed by Google for privacy-preserving telemetry~\cite{ErlingssonPK14}.
PI-RAPPOR uses pairwise independent hash functions to represent reports compactly, reducing communication from linear to logarithmic in the domain size.
With Hadamard Response, each client reports one index from an enlarged output domain, also requiring logarithmic communication.
Each HST client uses a public random vector of length $k$ and sends a single randomized bit.
Projective Geometry Response~\cite{FeldmanNNT22} also achieves optimal error guarantees, but implementing its sampler in our proof system requires non-native field arithmetic; Section~\ref{sec:protocol} discusses this overhead and our choice of verification mechanisms.

\subsubsection{RAPPOR and PI-RAPPOR}
\label{sec:prelim-rappor}
\label{sec:prelim-pi-rappor}

RAPPOR~\cite{ErlingssonPK14} and PI-RAPPOR~\cite{FeldmanT21} use different report formats but the same decoded-coordinate marginals.
Let $\alpha_0=(1+e^{\frac{\epsilon}{2}})^{-1}$ and $\alpha_1=1-\alpha_0$.
For each client with input $x\in \mathcal{X}$, both mechanisms encode bits $Z_0,\ldots,Z_{k-1}$ satisfying $\prob{Z_x=1\mid x}=\alpha_1$ and $\prob{Z_j=1\mid x}=\alpha_0$ for $j\neq x$.
The server decodes them as $\mathbf{y}=(y_0,\ldots,y_{k-1})\in\R^k$, where
\begin{equation}
    y_j=\frac{Z_j-\alpha_0}{\alpha_1-\alpha_0}
    \in\left\{-\frac{1}{e^{\frac{\epsilon}{2}}-1},
        \frac{e^{\frac{\epsilon}{2}}}{e^{\frac{\epsilon}{2}}-1}\right\}.
    \label{eq:rappor-report}
\end{equation}

With RAPPOR, the client samples each $Z_j$ independently by flipping $\mathbf{e}_x[j]$ with probability $\alpha_0$ and sends the bits $Z_0,\ldots,Z_{k-1}$ as its report.
Two one-hot inputs differ in two coordinates, so using randomized response with privacy parameter $\frac{\epsilon}{2}$ per coordinate gives $\epsilon$-LDP.

PI-RAPPOR instead encodes the bits in two elements of a prime-order field $\F_q$, where the field order $q$ upper bounds the number of classes $k$.
Assuming $\alpha_0q$ is an integer, define
\begin{equation}
    B(j)=\{(u,v)\in\F_q^2:
    \operatorname{int}(u+vj)<\alpha_0q\},
    \label{eq:pi-rappor-set}
\end{equation}
where $\operatorname{int}$ is the canonical representative in $\{0,\ldots,q-1\}$.
On input $x$, the client samples its report $Y\in\F_q^2$ uniformly from $B(x)$ with probability $\alpha_1$, or from its complement with probability $\alpha_0$.
Since $|B(x)|=\alpha_0q^2$,
\begin{equation}
    \prob{Y=y\mid x}=
    \begin{cases}
        \dfrac{e^{\frac{\epsilon}{2}}}{q^2},
            &y\in B(x),\\[2mm]
        \dfrac{e^{-\frac{\epsilon}{2}}}{q^2},
            &y\notin B(x).
    \end{cases}
    \label{eq:pi-ideal-pair-probability}
\end{equation}
Each probability changes by a factor of at most $e^\epsilon$ between inputs, so PI-RAPPOR is also $\epsilon$-LDP.
Decoding the pair as $Z_j\gets\case{1&\text{if}~Y\in B(j)\\0&\text{otherwise}}$ yields the stated marginals: membership in $B(x)$ has probability $\alpha_1$, while for $j\neq x$ the bijection $(u,v)\mapsto(u+vx,u+vj)$ gives probability $\alpha_0$.
In either mechanism, $\mathbb{E}[\mathbf{y}\mid x]=\mathbf{e}_x$.
Writing $\mathbf{y}_i$ for client $i$'s decoded vector, the server uses the same unbiased histogram estimator
\begin{equation}
    \widehat{F}=\sum_{i=0}^{n-1}\mathbf{y}_i.
    \label{eq:pi-ideal-estimator}
\end{equation}
For RAPPOR, computing this estimator takes $O\left(nk\right)$ time.
For PI-RAPPOR, as noted in~\cite{FeldmanNNT22}, dynamic programming with the chosen hash function yields an aggregation time of $O\left(n+k\log k\right)$.

\subsubsection{Hadamard Response~\cite{AcharyaSZ19}}
\label{sec:prelim-hadamard}
This mechanism makes use of the Sylvester Hadamard matrix, defined as
\begin{equation}
    H_K=
    \begin{bmatrix}
        1 & 1\\
        1 & -1
    \end{bmatrix}^{\otimes\log_2 K}
    \in\{-1,1\}^{K\times K}
\end{equation}
where $K=2^{\lceil\log_2(k+1)\rceil}$.
We associate input $x\in\mathcal{X}$ with row $x+1$, thereby excluding the all-one row.
With Hadamard Response, the client samples a column index $y\in\{0,\ldots,K-1\}$ as its report according to
\begin{equation}
    \prob{\mathcal{M}_{\mathsf{HR}}(x)=y}
    =\begin{cases}
        \dfrac{2e^\epsilon}{K(e^\epsilon+1)},
            & H_K[x+1,y]=1,\\[2mm]
        \dfrac{2}{K(e^\epsilon+1)},
            & H_K[x+1,y]=-1.
    \end{cases}
    \label{eq:hadamard-response}
\end{equation}
The likelihood ratio between any two inputs is at most $e^\epsilon$, so the mechanism is $\epsilon$-LDP.
Given reports $y_0,\ldots,y_{n-1}$, the server uses
\begin{equation}
    \widehat{F}_x=
    \frac{e^\epsilon+1}{e^\epsilon-1}
    \sum_{i=0}^{n-1} H_K[x+1,y_i]
    \label{eq:hadamard-estimator}
\end{equation}
for each $x\in\mathcal{X}$.
Orthogonality of the Hadamard rows gives $\mathbb{E}[\widehat{F}_x]=F(\mathbf{x})_x$.
Hadamard Response is asymptotically optimal for discrete distribution estimation in the high privacy regime and requires only $\log_2 K$ communicated bits per report.
Using fast Hadamard matrix multiplication, one can compute the estimator in $O\left(n+k\log k\right)$ time.

\subsubsection{HST}
\label{sec:prelim-hst}

HST is a public-randomness histogram protocol described by Bassily et al.~\cite{BassilyNST20}.
For each client $i$ with input $x_i\in \mathcal{X}$, the protocol independently samples a public vector $\mathbf{s}_i\in\{-1,1\}^k$ uniformly at random.
Client $i$ applies binary randomized response to the $x_i$th bit of the public vector, $s_{i,x_i}$, sending
\begin{equation}
    y_i=\begin{cases}
        s_{i,x_i},&\text{with probability }\dfrac{e^\epsilon}{e^\epsilon+1},\\
        -s_{i,x_i},&\text{with probability }\dfrac{1}{e^\epsilon+1}.
    \end{cases}
\end{equation}
Write $\lambda_\epsilon=\frac{e^\epsilon+1}{e^\epsilon-1}$ and decode the bit as $\mathbf{y}_i=\lambda_\epsilon y_i\mathbf{s}_i$.
The count estimator is
\begin{equation}
    \widehat F=\sum_{i=0}^{n-1}\mathbf{y}_i.
    \label{eq:hst-estimator}
\end{equation}
The public vectors are fixed by the protocol, not chosen by the reporting clients.
Conditional on these vectors, binary randomized response is $\epsilon$-LDP; averaging also over the public randomness gives $\mathbb{E}[\mathbf{y}_i\mid x_i]=\mathbf{e}_{x_i}$.
HST attains an optimal error rate when all clients follow the protocol~\cite{YeB18}.
In this explicit form it uses $nk$ public random bits and $O\left(nk\right)$ aggregation time, although each client sends only one bit.

\subsection{Zero-Knowledge Proofs of Knowledge}
\label{sec:prelim-crypto-zkp}

\newcommand{\privin}{x_\text{priv}}
\newcommand{\pubin}{x_\text{pub}}

A commitment scheme allows a sender to bind itself to a value without revealing that value.
We write $c\gets\mathsf{Commit}(x;\omega)$ for a commitment to $x$, where $\omega$ is the commitment randomness used to hide $x$.
We require the scheme to be computationally \emph{binding}, meaning that no probabilistic polynomial-time (PPT) sender can open one commitment to two different values, and
\emph{hiding}, meaning that commitments to any two values are
computationally indistinguishable.
These properties let the protocol refer to a fixed client value while keeping that value private~\cite{Thaler22}.

We use zero-knowledge proofs of knowledge (ZKPoKs) to establish that a committed computation was performed correctly~\cite{BellareG92, GoldwasserMR85, Thaler22}.
For a deterministic function $f$ that takes public input  $\pubin$ and private input $\privin$ and outputs $y$, let $z=(\pubin,y,c)$ be the public
\emph{statement} and $w=(\privin,\omega)$ be the private \emph{witness}. Define
the relation
\begin{equation}
    \begin{aligned}
        \mathcal{R}_f=\bigl\{(z,w):\;&c=\mathsf{Commit}(\privin;\omega)\\
        &\land\ y=f(\privin,\pubin)\bigr\}.
    \end{aligned}
\end{equation}
A ZKPoK for $\mathcal{R}_f$ is a proof protocol
\begin{equation}
    \Pi(z,w):=\left\langle\prover(z,w),\verifier(z)\right\rangle,
    \label{eq:zkp}
\end{equation}
in which the prover $\prover$ convinces the verifier $\verifier$ that it knows a witness $w$ such that $(z,w)\in\mathcal{R}_f$, without revealing the witness.
For security parameter $\kappa\in\N$, we use the following standard properties:
\begin{itemize}[leftmargin=*]
    \item \textbf{Completeness.} If $(z,w)\in\mathcal{R}_f$ and both parties
    follow $\Pi$, then $\verifier$ accepts except with negligible probability in $\kappa$.
    \item \textbf{Knowledge soundness.} For any PPT prover that makes
    $\verifier$ accept with non-negligible probability, acceptance guarantees that the prover knows a witness $w$ satisfying $(z,w)\in\mathcal{R}_f$.
    \item \textbf{Zero knowledge.} The verifier learns nothing about the
    private witness $w$ beyond the validity of the public statement $z$.
\end{itemize}

In our protocols, the witness contains both the client's input and its private randomness share.
The client first proves knowledge of a valid commitment opening before receiving the server's randomness share.
It later proves that the submitted report was computed using an opening of the same commitment.
Together with commitment binding, these proofs tie every accepted report to the client's input and randomness share fixed before the server's share was revealed.

\subsubsection{Verifiable Randomized Computation via Distributed Randomness} \label{sec:distributed-randomness}
A randomized mechanism $\mathcal{M}(x)$ can be represented by a deterministic function $f(x,r^{\cli},r^{\ser})$, where $r^{\cli}$ and $r^{\ser}$ are randomness shares contributed by the client and server, respectively.
\begin{Def}[Distributed randomness for an LDP mechanism]
    \label{def:distributed-randomness}
    A deterministic function $f$ and prescribed distributions for the client and server randomness shares implement an $\epsilon$-LDP mechanism $\mathcal{M}$ through distributed randomness if they satisfy the following requirements:
    \begin{enumerate}[leftmargin=*]
        \item \textbf{(Integrity)} For every fixed input $x$ and valid client share $r^{\cli}$, sampling the server's share $r^{\ser}$ from its prescribed distribution makes $f(x,r^{\cli},r^{\ser})$ computationally indistinguishable from $\mathcal{M}(x)$.
        \item \textbf{(Privacy)} When the client samples its share honestly and independently of the server's share, the joint view $(r^{\ser},f(x,r^{\cli},r^{\ser}))$ satisfies $\epsilon$-computational LDP with respect to $x$.
    \end{enumerate}
\end{Def}
Given integrity, a sufficient condition for privacy is that $f(x,r^{\cli},r^{\ser})$ and $r^{\ser}$ are independent for every fixed $x$ when the shares are sampled honestly and independently.
Indeed, integrity ensures that the marginal distribution of the output is computationally indistinguishable from $\mathcal{M}(x)$, and independence ensures that revealing $r^{\ser}$ does not affect its privacy.
Our instantiations satisfy this condition.
The server's contribution prevents a malicious client from selecting all of the random coins and biasing its report; the client's contribution prevents the server from determining those coins by itself and learning $x$~\cite{BiswasC23, ChampionSU19}.
The client then proves the relation above with private input $\privin=(x,r^{\cli})$, public input $\pubin=r^{\ser}$, and public output $y\gets f(x,r^{\cli},r^{\ser})$.

\section{Problem Setup}\label{sec:setup}

We consider one server and $n$ clients indexed by $0,\ldots,n-1$.
Each client $i$ holds one categorical value $x_i\in\mathcal{X}$, and the server seeks to estimate the histogram $F(\mathbf{x})$ from reports generated with privacy budget $\epsilon$.
The server acts as the verifier in the cryptographic protocols.
It is semi-honest: it follows the prescribed protocol but may attempt to infer additional information from its view.
Without loss of generality, the first $m$ clients are corrupted, with $C=\{0,\ldots,m-1\}$.
An adversary may coordinate the corrupted clients, which can deviate arbitrarily from the client algorithm; the remaining clients follow the protocol with independent randomness.
In particular, the adversary may coordinate the corrupted clients' reports through output manipulation, as defined in Definition~\ref{def:output-manipulation}.

\begin{Def}[Output manipulation]
	\label{def:output-manipulation}
	The adversary may jointly choose arbitrary reports $\widetilde y_i\in\mathcal{Y}$ for the corrupted clients $i\in C$, where $\mathcal{Y}$ is the report space of the local randomizer $\mathcal{M}$.
	These reports need not follow the distribution prescribed by $\mathcal{M}$.
\end{Def}

Our goal is twofold: the server's view of each honest client must reveal no more than the prescribed $\epsilon$-LDP guarantee, and every accepted report must be consistent with some committed input and the prescribed randomization procedure.
However, consistency with a valid execution alone does not prevent manipulation of the submitted report distribution: a client may try multiple valid executions and submit only a favorable report.
Section~\ref{sec:ideal-impossibility} examines this resampling attack in one-round-trip designs with client-visible reports.
We then introduce a commitment step that fixes the client's input and randomness before the server reveals its share, preventing resampling while still allowing selective dropout.

\subsection{Limitations of One-Round-Trip Protocols}
\label{sec:ideal-impossibility}

An appealing approach is to verify an LDP report in a single round trip: the server sends one challenge, the client returns one response based on the challenge and input data, and the server verifies the response before using the report to estimate the histogram.
We use $y=f(x,r^{\cli},r^{\ser})\in\mathcal{Y}$ to denote one client's randomized report, computed from its input $x$ and the client and server randomness shares $r^{\cli}$ and $r^{\ser}$.
The client's response is the full protocol message carrying this report, either in plaintext or encrypted, together with the required proof.
The server uses accepted reports to compute the histogram estimate.
This communication-efficient server--client--server pattern is used in V$\epsilon$rity~\cite{BellClarkGLRC25}.
However, if the client can compute the report from its local view, it may try multiple executions locally for the same server challenge and submit only a favorable report with its proof, enabling \emph{output manipulation}.
We first analyze this attack, then explain how output hiding via encryption prevents it, and finally discuss alternative design choices that allow lightweight plaintext evaluation.

\subsubsection{Why Output Visibility Enables Output Manipulation}

In a one-round-trip server--client--server protocol, a client that can compute its report from its local view may repeatedly sample client randomness for the same input and server challenge.
It can then submit a favorable report together with a valid proof for the selected execution.
Each candidate is computed correctly, but selection changes the submitted report distribution.
The following theorem formalizes this resampling attack when favorable reports occur with inverse-polynomial probability.

\begin{Thm}[Output manipulation from client-visible outputs]
    \label{thm:client-predictable-bias}
    Let $\Pi$ be a correct PPT one-round-trip server--client--server implementation of a randomized mechanism $\mathcal{M}$.
    For a fixed input $x\in\mathcal{X}$, write $r^{\ser}$ for the server's challenge, $r^{\cli}$ for the client's randomness, and $y=f(x,r^{\cli},r^{\ser})\in\mathcal{Y}$ for the randomized report.
    Assume there are a PPT algorithm $\mathsf{ComputeReport}$, an efficiently recognizable set $S\subseteq\mathcal{Y}$, and a polynomial $p$ such that, except with negligible probability over the server's challenge, $\mathsf{ComputeReport}$ computes $y$ from the client's local view with negligible error and
    \begin{equation}
        \frac{1}{p(\kappa)}
        \leq
        \prob[r^{\cli}]{y\in S\mid x,r^{\ser}}
        \leq \frac{1}{2}.
        \label{eq:predictable-event}
    \end{equation}
    Then a PPT client can submit a valid response that the server accepts with report $y\in S$ with probability $1-\mathsf{negl}(\kappa)$.
    The distribution of the submitted report has total variation distance at least $\frac{1}{2}-\mathsf{negl}(\kappa)$ from the honest report distribution.
\end{Thm}

\begin{proof}[Proof sketch]
    Rejection sampling finds a favorable report in polynomial time with overwhelming probability.
    See Appendix~\ref{app:client-predictable-bias} for the proof.
\end{proof}

\subsubsection{Hiding Reports with Encryption and Its Cost}

Output visibility also enables a distinct attack: even when restricted to a single prescribed report, a client may withhold it after inspection, resulting in selective dropout (Definition~\ref{def:selective-dropout}).

\begin{Def}[Selective dropout]
	\label{def:selective-dropout}
	Each corrupted client $i\in C$ has one report $y_i$ generated by a single execution of the prescribed randomizer on its fixed input $x_i$.
	After inspecting these reports, the adversary may choose a subset $D\subseteq C$ of clients to drop out; each client in $C\setminus D$ submits its original report $y_i$ unchanged.
	The adversary may coordinate this choice across corrupted clients, but may not modify the reports or repeat their randomization.
\end{Def}

Hiding candidate reports throughout the client's computation prevents selection based on their realized values, addressing both resampling and selective dropout. 
However, to achieve this while retaining a single round trip, the protocol must let the server verify the report without revealing it to the client during computation.
The client therefore cannot simply evaluate the mechanism in plaintext and send a proof of the result.
A natural approach is to evaluate the mechanism on encrypted data and use zero-knowledge proofs to certify the required cryptographic operations.
V$\epsilon$rity follows this approach: the server encrypts its randomness, the client computes an encryption of its report using HE, and zero-knowledge proofs certify the required HE operations~\cite{BellClarkGLRC25}.
The server decrypts the report after verifying the proofs of the HE operations.

This design incurs the cost of computation using HE and of proving HE operations.
The proofs certify the arithmetic of encryption and partial decryption, rather than just the comparatively small plaintext LDP computation.
Proving these cryptographic operations compounds the cost of HE itself: the client must not only perform expensive cryptographic operations but also prove their correctness.
This makes report generation substantially more costly than evaluating and proving the plaintext LDP mechanism.
For binary randomized response, V$\epsilon$rity reports 1,423.50\,s of online client computation and 1,472.77\,s of server computation, approximately 23.7 and 24.5 minutes, respectively~\cite[Section~9.2]{BellClarkGLRC25}.

\subsection{Lightweight Design Choices}

Although output hiding prevents selection based on the realized report, the cost of this encrypted construction makes it impractical for large-scale telemetry.
We therefore pursue two alternatives, depending on whether the client protocol can be changed.

First, our lightweight construction adds a commitment message before the server reveals its randomness share.
The commitment fixes the client's input and randomness share before the server supplies its share, which then determines a single report that can pass verification.
The client can then compute its report in plaintext and prove correct evaluation without HE.
It cannot resample this fixed report, but it can still decide whether to submit it after inspecting it, enabling selective dropout.

\textbf{Section~\ref{sec:lightweight-protocol}} presents the framework, followed by concrete instantiations in \textbf{Section~\ref{sec:concrete-instantiations}}.
Although malicious clients may still withhold their reports after inspecting them, \textbf{Section~\ref{sec:dropout-analysis}} shows that, for (PI-)RAPPOR and Hadamard Response, this selective dropout causes asymptotically less attack-dependent error than output manipulation.
The extra message therefore lets us avoid the overhead caused by HE while retaining a meaningful robustness benefit.
Rather than hiding every report from its client, we enforce correct randomization and quantify the remaining error from selective dropout.

Second, when adding commitments and proofs to the client protocol is not feasible, the server can instead check the statistical consistency of the collected reports.
\textbf{Section~\ref{sec:cons}} develops such post-hoc checks for RAPPOR and Hadamard Response.
These checks do not certify each client's computation, but they can limit the aggregate damage from output manipulation without any client-side changes.

\section{An Efficient and Secure Construction of Verifiable LDP Mechanisms}\label{sec:protocol}

We construct efficient verifiable LDP protocols that prevent output manipulation while allowing selective dropout.
As we show in Section~\ref{sec:dropout-analysis}, for (PI-)RAPPOR and Hadamard Response, the attack-dependent error from selective dropout is smaller by a factor of $\sqrt{k}$ than that from output manipulation, where $k$ is the domain size.
Thus, verification can improve robustness without preventing dropout.

For our implementations, we choose PI-RAPPOR and Hadamard Response because their report-generation computations are sublinear in $k$ and admit compact circuits over the fixed prime field used by our ZKPoK backend.
PI-RAPPOR generates a report consisting of two field elements, while Hadamard Response generates an index represented by $O\left(\log k\right)$ bits.
In contrast, directly verifying RAPPOR requires generating and proving the randomization of $k$ report bits.
For HST, restricting output manipulation to selective dropout does not improve the worst-case asymptotic error rate, as shown in Section~\ref{sec:dropout-analysis}.
Accordingly, HST serves as a comparison mechanism in our analysis and experiments.

Projective Geometry Response~\cite{FeldmanNNT22} is another frequency-estimation mechanism with optimal error guarantees, but its sampler uses a finite field different from the backend's native field.
Implementing its field operations would require repeated non-native modular arithmetic, including modular reduction and range constraints, adding circuit overhead.
We leave its efficient verification to future work.

We first present a general three-message framework that fixes the client's input and randomness before the server supplies its randomness share.
We then instantiate it for PI-RAPPOR and Hadamard Response, using plaintext evaluation and compact zero-knowledge proofs to enforce correct randomization without HE.

\subsection{A Framework for Verifiable LDP Against Output Manipulation}
\label{sec:lightweight-protocol}

Our protocol uses one additional client message to enable substantially cheaper plaintext computation and proofs.
The key ordering constraint is that the client commits to both its input and its randomness share before learning the server's randomness share.
For that fixed commitment, the server's share determines the only report that the client can prove valid.
Thus, a malicious client can submit that report or submit nothing, but it cannot choose among multiple valid reports for the same commitment after seeing the server's randomness share.

Let $f(x,r^{\cli},r^{\ser})$ be the deterministic realization of an $\epsilon$-LDP mechanism described in Section~\ref{sec:distributed-randomness}.
The protocol uses a commitment scheme $\mathsf{Commit}$ and two ZKPoKs, with the statements specified in the steps below.

For each client $i$, the protocol proceeds as follows.

\begin{enumerate}[leftmargin=*]
	\item \textbf{Commitment:} Client $i$ samples a private randomness share $r_i^{\cli}$ and commitment randomness $\omega_i$, and computes $c_i=\mathsf{Commit}(x_i\mathbin\|r_i^{\cli};\omega_i)$, where $\mathbin\|$ denotes concatenation.
		It sends $c_i$ together with a ZKPoK $\pi_i^{\mathsf{in}}$ of knowledge of $(x_i,r_i^{\cli},\omega_i)$ such that $x_i\in\mathcal{X}$, $r_i^{\cli}$ satisfies the mechanism's validity constraints, and
	\begin{equation}
		c_i=\mathsf{Commit}(x_i\mathbin\|r_i^{\cli};\omega_i).
		\label{eq:input-commitment-relation}
	\end{equation}
    The server verifies the proof.
    A client enters the protocol only if this first proof is accepted; otherwise, the client is considered to have exited before the server generates its randomness share.
	\item \textbf{Server randomness:} For every admitted client $i$, the server independently samples a randomness share
	$r_i^{\ser}$ from the distribution prescribed by the distributed-randomness construction and sends it to that client.
	Sampling $r_i^{\ser}$ only after accepting $c_i$ ensures that the committed pair $(x_i,r_i^{\cli})$ cannot depend on the server's share.
    \item \textbf{Report and verification:} Client $i$ computes
    \begin{equation}
			y_i=f(x_i,r_i^{\cli},r_i^{\ser})
    \end{equation}
	and sends $y_i$ together with a ZKPoK $\pi_i^{\mathsf{eval}}$ of knowledge of $(x_i,r_i^{\cli},\omega_i)$ satisfying
	\begin{equation}
		\begin{aligned}
			c_i&=\mathsf{Commit}(x_i\mathbin\|r_i^{\cli};\omega_i),\\
			y_i&=f(x_i,r_i^{\cli},r_i^{\ser}).
		\end{aligned}
		\label{eq:evaluation-relation}
	\end{equation}
    The server accepts $y_i$ if and only if this proof verifies with respect to $(c_i,r_i^{\ser},y_i)$.
    It then aggregates all accepted reports using the estimator associated with the underlying LDP mechanism.
\end{enumerate}

\begin{Thm}[Protocol guarantees]
	\label{thm:lightweight-protocol}
	Assume that the commitment scheme is binding and hiding, both ZKPoKs are complete, knowledge sound, and zero knowledge, and the distributed randomness construction satisfies Definition~\ref{def:distributed-randomness}.
	Then the protocol above has the following properties:
	\begin{enumerate}[leftmargin=*]
		\item \textbf{Completeness.} Every honest client that remains online
		is accepted except with negligible probability.
		\item \textbf{Accepted-output integrity.} Except with negligible
		probability, every accepted report $y_i$ is equal to $f(x_i,r_i^{\cli},r_i^{\ser})$ for the input and client randomness share fixed by $c_i$ before $r_i^{\ser}$ was sampled.
		\item \textbf{Computational privacy.} For an honest client, the
		server's complete view satisfies $\epsilon$-computational local differential privacy as in Definition~\ref{def:computational-ldp}.
	\end{enumerate}
\end{Thm}

\begin{proof}[Proof sketch]
	The guarantees follow from the commitment, ZKPoK, and distributed-randomness properties.
	See Appendix~\ref{app:lightweight-protocol} for the proof.
\end{proof}

\subsection{Circuit Instantiations for Frequency Estimation}
\label{sec:concrete-instantiations}

We instantiate the framework of Section~\ref{sec:lightweight-protocol} for PI-RAPPOR and Hadamard Response by expressing their samplers as deterministic circuits over client and server randomness shares.
The constructions show how to combine these shares and implement the sampling probabilities at finite precision while preserving $\epsilon$-LDP.
For either mechanism, the combined randomness has the prescribed distribution whenever either party's shares are sampled honestly and independently of the other party's shares.

The circuits below compute $f(x,r^{\cli},r^{\ser})$, whose evaluation is certified in the final protocol step.
We use $\gamma$-bit integer thresholds for probabilistic choices, with conservative rounding to preserve the privacy guarantee.
The proofs additionally enforce input ranges, bitness, commitment openings, and consistency with the public outputs; we omit these constraints from the pseudocode.

\subsubsection{PI-RAPPOR Circuit}
Algorithm~\ref{alg:pi-rappor-circuit} implements the PI-RAPPOR sampler from Section~\ref{sec:prelim-pi-rappor} using distributed randomness.
The prime $q\geq k$ is the order of the field $\F_q$ in which the two report elements are represented; $\gamma$ is the number of bits used for a random branch choice.
Recall that $\alpha_0=\frac{1}{1+e^{\frac{\epsilon}{2}}}$ is the probability that a nonmatching coordinate is encoded as one.
We set $\tau_{\mathsf{PI}}=\lceil q\alpha_0\rceil$ and define the rounded version of the set in Equation~\eqref{eq:pi-rappor-set} by $B_\tau(j)=\{(u,v)\in\F_q^2:\operatorname{int}(u+vj)<\tau_{\mathsf{PI}}\}$.
We also set $\theta_{\mathsf{PI}}=\lfloor2^\gamma(1-e^{-\frac{\epsilon}{2}})\rfloor$; this threshold determines how often the sampler selects a pair from $B_\tau(x)$ instead of a uniform pair from $\F_q^2$.

For each party $P\in\{\cli,\ser\}$, an honestly sampled randomness share contains independent uniform values $r_0^P,r_1^P\in\F_q$, an integer $z^P\in\{0,\ldots,\tau_{\mathsf{PI}}-1\}$, and a bit string $d^P\in\{0,1\}^\gamma$.
The circuit combines the two parties' values to obtain the sampler's random choices.
The combined value $r_1$ is the second component of the report in both branches, while $r_0$ is the first component in the uniform-pair branch.
In the restricted-pair branch, $z$ and $r_1$ instead determine a pair in $B_\tau(x)$.
The combined bit strings select which branch to use.
The circuit proceeds as follows:
\begin{itemize}[leftmargin=*]
	\item Lines~\ref{algline:pi-mask-zero} and~\ref{algline:pi-mask-one} add $r_0^{\cli}$ to $r_0^{\ser}$ and $r_1^{\cli}$ to $r_1^{\ser}$, respectively, using addition in $\F_q$.
	The resulting $r_0$ and $r_1$ are independent and uniform in $\F_q$ when either party samples its share honestly and independently of the other's share.
	\item Line~\ref{algline:pi-modular-share} adds $z^{\cli}$ and $z^{\ser}$ modulo $\tau_{\mathsf{PI}}$ to obtain a uniform $z\in\{0,\ldots,\tau_{\mathsf{PI}}-1\}$.
	In the restricted-pair branch, $z$ will equal the canonical representative of $y_0+y_1x$ and thus ensure that the report is in $B_\tau(x)$.
	\item Line~\ref{algline:pi-branch-value} XORs $d^{\cli}$ and $d^{\ser}$ and interprets the resulting bits as a uniform integer $U\in\{0,\ldots,2^\gamma-1\}$.
	Comparing $U$ with $\theta_{\mathsf{PI}}$ lets the circuit select the restricted-pair branch with probability $\frac{\theta_{\mathsf{PI}}}{2^\gamma}$.
	\item Line~\ref{algline:pi-branch} sets $b=0$ when $U<\theta_{\mathsf{PI}}$, selecting the restricted-pair branch, and $b=1$ otherwise, selecting the uniform-pair branch.
	\item Lines~\ref{algline:pi-output-zero} and~\ref{algline:pi-output-one} construct the report.
	When $b=0$, the report is $(z-r_1x,r_1)$, so $\operatorname{int}(y_0+y_1x)=z<\tau_{\mathsf{PI}}$ and the pair belongs to $B_\tau(x)$.
	Each pair in $B_\tau(x)$ corresponds to exactly one choice of $z$ and $r_1$, so uniform, independent choices of these values give a uniform pair in $B_\tau(x)$.
	When $b=1$, the report is $(r_0,r_1)$, a uniform pair in $\F_q^2$.
	The circuit returns the resulting pair $y=(y_0,y_1)$.
\end{itemize}

Write $\widetilde\delta=\frac{\theta_{\mathsf{PI}}}{2^\gamma}$ for the probability of choosing the restricted-pair branch, $\widetilde\alpha_0=\frac{\tau_{\mathsf{PI}}}{q}$, and $\widetilde\alpha_1=\widetilde\delta+(1-\widetilde\delta)\widetilde\alpha_0$.
With the prescribed randomness shares, the client obtains a uniform pair from $B_\tau(x)$ with probability $\widetilde\delta$ and a uniform pair from $\F_q^2$ otherwise; the latter branch can also produce a pair in $B_\tau(x)$.
The resulting probabilities of membership in $B_\tau(j)$ are $\widetilde\alpha_1$ for $j=x$ and $\widetilde\alpha_0$ for $j\ne x$, corresponding to the matching and nonmatching probabilities in Section~\ref{sec:prelim-pi-rappor}.
The largest likelihood ratio is at most $1+\frac{\widetilde\delta}{\widetilde\alpha_0(1-\widetilde\delta)}\leq e^\epsilon$, so the rounded mechanism satisfies $\epsilon$-LDP.
Rounding changes the nonmatching and matching coordinate probabilities by less than $\frac{1}{q}$ and $\frac{1}{q}+2^{-\gamma}$, respectively; when $\alpha_0q\in\N$, the full report distribution is within $2^{-\gamma}$ in total variation of PI-RAPPOR.
The estimator in Equation~\eqref{eq:pi-ideal-estimator} remains unbiased when membership in $B_\tau(j)$ and the probabilities $\widetilde\alpha_0,\widetilde\alpha_1$ are used in Equation~\eqref{eq:rappor-report}; we choose $\gamma$ so that $\widetilde\delta>0$.

\begin{algorithm}[t]
	\caption{PI-RAPPOR circuit}
	\label{alg:pi-rappor-circuit}
	\autofit{\begin{algorithmic}[1]
		\Require Class $x\in\{0,\ldots,k-1\}$
		\Require Parameters $(q,\gamma,\tau_{\mathsf{PI}},\theta_{\mathsf{PI}})$
		\Require Client share $r^{\cli}=(r_0^{\cli},r_1^{\cli},z^{\cli},d^{\cli})$
		\Require Server share $r^{\ser}=(r_0^{\ser},r_1^{\ser},z^{\ser},d^{\ser})$
		\Require $r_0^\cli,r_1^\cli\in\F_q$,
		$z^\cli\in\{0,\ldots,\tau_{\mathsf{PI}}-1\}$, $d^\cli\in\{0,1\}^\gamma$
        \Require $r_0^\ser,r_1^\ser\in\F_q$,
		$z^\ser\in\{0,\ldots,\tau_{\mathsf{PI}}-1\}$, $d^\ser\in\{0,1\}^\gamma$
		\Ensure Compact report $y=(y_0,y_1)\in\F_q^2$
		\State $r_0\gets r_0^{\cli}+r_0^{\ser}$ in $\F_q$ \Comment{Combine the randomness shares}
		\label{algline:pi-mask-zero}
		\State $r_1\gets r_1^{\cli}+r_1^{\ser}$ in $\F_q$
		\label{algline:pi-mask-one}
		\State $z\gets(z^{\cli}+z^{\ser})\bmod\tau_{\mathsf{PI}}$
		\label{algline:pi-modular-share}
		\State $U\gets\sum_{i=0}^{\gamma-1}2^i(d_i^{\cli}\oplus d_i^{\ser})$
		\label{algline:pi-branch-value} \Comment{Choose the mixture component}
		\State $b\gets\mathbbm{1}[U\geq\theta_{\mathsf{PI}}]$
		\label{algline:pi-branch} \Comment{Uniform pair if $b=1$; restricted pair otherwise}
		\State $y_0\gets(1-b)(z-r_1x)+br_0$
		\label{algline:pi-output-zero}
		\State $y_1\gets r_1$
		\label{algline:pi-output-one}
		\State \Return $y=(y_0,y_1)$
	\end{algorithmic}}
\end{algorithm}

\subsubsection{Hadamard Response Circuit}
Algorithm~\ref{alg:hr-circuit} implements the Hadamard Response sampler from Section~\ref{sec:prelim-hadamard} using distributed randomness.
Recall that $K=2^{\lceil\log_2(k+1)\rceil}$, input $x$ corresponds to row $x+1$ of $H_K$, and the report is a column index $y$.
Let $\ell=\log_2K$, so both the row and column indices have $\ell$-bit representations.
We write $\mathsf{Bits}_{\ell}(t)$ for the $\ell$-bit representation of an integer $t$, with the least significant bit indexed by zero.
In Equation~\eqref{eq:hadamard-response}, the report is chosen from the columns with $H_K[x+1,y]=-1$ with total probability $\frac{1}{1+e^\epsilon}$ and from those with $H_K[x+1,y]=1$ otherwise.
The circuit uses $\gamma$ bits to choose between these two sets and rounds the threshold up to $\tau_{\mathsf{HR}}=\lceil\frac{2^\gamma}{1+e^\epsilon}\rceil$.

Each party $P\in\{\cli,\ser\}$ contributes a randomness share containing an $\ell$-bit string $c^P$ for choosing a column and a $\gamma$-bit string $a^P$ for choosing the sign of the Hadamard entry.
These strings are combined to form the candidate column $c$ and the scalar parity choice $a$ below.
For bit representations of a row and column, the Hadamard entry is $(-1)$ raised to the parity of their bitwise inner product.
The circuit chooses the sign, then adjusts one column bit so that this parity agrees with the chosen sign:
\begin{itemize}[leftmargin=*]
	\item Line~\ref{algline:hr-input-bits} writes the row index $x+1$ as an $\ell$-bit string $b$, least significant bit first.
	Since $x+1\ne0$, at least one bit of $b$ is set.
	\item Line~\ref{algline:hr-column-share} XORs the parties' column strings to obtain a uniformly random candidate column $c$.
	\item Line~\ref{algline:hr-branch-value} XORs their $\gamma$-bit strings and interprets the result as an integer $U\in\{0,\ldots,2^\gamma-1\}$.
	\item Line~\ref{algline:hr-branch} sets $a=1$ when $U<\tau_{\mathsf{HR}}$ and $a=0$ otherwise.
	The value $a$ is the desired parity: $a=1$ selects an entry of $-1$, and $a=0$ selects an entry of $1$.
	\item Line~\ref{algline:hr-mask} chooses $j^\star$, the index of the least-significant set bit of $b$, as the pivot.
	\item Line~\ref{algline:hr-parity} computes $\rho$, the parity contributed by all row and column bits except the pivot.
	\item Line~\ref{algline:hr-replace-bit} sets the pivot bit of the output column to $a\oplus\rho$ and leaves its other bits unchanged.
	Because $b_{j^\star}=1$, the complete bitwise inner product has parity $\rho\oplus(a\oplus\rho)=a$.
	\item Line~\ref{algline:hr-output} converts the resulting column bits to the integer report $y$ and returns it.
\end{itemize}

The unchanged column bits are uniform, so the resulting report is uniform among the columns with the selected sign.
The circuit selects the $-1$ class with probability $\widetilde\alpha_{\mathsf{HR}}=\frac{\tau_{\mathsf{HR}}}{2^\gamma}$, which differs from $\frac{1}{1+e^\epsilon}$ by less than $2^{-\gamma}$.
The report distribution therefore differs from Hadamard Response by less than $2^{-\gamma}$ in total variation.
Rounding the threshold up gives $\frac{1-\widetilde\alpha_{\mathsf{HR}}}{\widetilde\alpha_{\mathsf{HR}}}\leq e^\epsilon$, so the rounded mechanism satisfies $\epsilon$-LDP.
Replacing the coefficient in Equation~\eqref{eq:hadamard-estimator} by $\frac{1}{1-2\widetilde\alpha_{\mathsf{HR}}}$ yields an unbiased estimator; we choose $\gamma$ so that $\widetilde\alpha_{\mathsf{HR}}<\frac{1}{2}$.

\begin{algorithm}[t]
	\caption{Hadamard Response circuit}
	\label{alg:hr-circuit}
	\autofit{\begin{algorithmic}[1]
		\Require Class $x\in\{0,\ldots,k-1\}$
		\Require Parameters $(\ell,\gamma,\tau_{\mathsf{HR}})$, with $K=2^\ell$
		\Require Client share $r^{\cli}=(c^{\cli},a^{\cli})$
		\Require Server share $r^{\ser}=(c^{\ser},a^{\ser})$
		\Require For each $P\in\{\cli,\ser\}$:
		$c^P\in\{0,1\}^\ell$, $a^P\in\{0,1\}^\gamma$
		\Ensure Hadamard report $y\in\{0,\ldots,K-1\}$
		\State $b\gets\mathsf{Bits}_{\ell}(x+1)$
		\label{algline:hr-input-bits}
		\Comment{least significant bit first}
		\State $c\gets c^{\cli}\oplus c^{\ser}$ \Comment{Combine the randomness shares}
		\label{algline:hr-column-share}
		\State $U\gets\sum_{i=0}^{\gamma-1}2^i(a_i^{\cli}\oplus a_i^{\ser})$
		\label{algline:hr-branch-value} \Comment{Choose the output parity}
		\State $a\gets\mathbbm{1}[U<\tau_{\mathsf{HR}}]$
		\label{algline:hr-branch} \Comment{Select a pivot and collect the remaining parity}
		\State $j^\star\gets\min\{i:b_i=1\}$
		\label{algline:hr-mask}
		\State $\rho\gets\bigoplus_{\substack{0\leq i<\ell\\i\neq j^\star}}b_i c_i$
		\label{algline:hr-parity} \Comment{Set the pivot bit to enforce parity $a$}
		\State $c'\gets c$; $c'_{j^\star}\gets a\oplus\rho$
		\label{algline:hr-replace-bit}
		\State \Return $y\gets\sum_{i=0}^{\ell-1}2^ic'_i$
		\label{algline:hr-output}
	\end{algorithmic}}
\end{algorithm}

\section{Analysis of Selective Dropout Attacks}\label{sec:dropout-analysis}

The protocol in Section~\ref{sec:protocol} prevents output manipulation but leaves malicious clients free to withhold their prescribed reports.
We now quantify the error caused by this selective dropout and compare it with the error under output manipulation.
For (PI-)RAPPOR and Hadamard Response, the attack-dependent error from selective dropout is smaller by a factor of $\sqrt{k}$ than under output manipulation.
For HST, selective dropout and output manipulation have the same asymptotic effect.
This comparison establishes when the restriction enforced by the protocol improves robustness and explains the choice of PI-RAPPOR and Hadamard Response for our instantiations.
Thus, verification can reduce the damage from malicious clients even when selective dropout remains possible.

As in Section~\ref{sec:setup}, $C=\{0,\ldots,m-1\}$ is the set of corrupted clients.
Client $i\in C$ holds a committed input $x_i$ and generates an independently randomized valid report $y_i$.
After observing all corrupted clients' reports, the adversary chooses a subset $D\subseteq C$ of clients to drop out.
The server applies the mechanism's usual estimator to the reports from $\{0,\ldots,n-1\}\setminus D$.
We measure its expected $\ell_1$ error against the histogram $F(\mathbf{x})$ of the $n$ admitted clients' committed inputs, including those of dropped clients:
\begin{equation}
	\mathsf{Err}_1
	=\mathbb{E}\left[\left\|\widehat{F}-F(\mathbf{x})\right\|_1\right].
	\label{eq:dropout-error}
\end{equation}
The expectation is over mechanism randomness and any internal randomness of the attack; the choice of $D$ may be an arbitrary randomized function of the corrupted clients' reports.
For each mechanism, let $\mathbf{v}_i$ be the contribution of client $i$'s report to its usual histogram estimator, including any decoding and scaling.
If $\widehat{F}_{\mathsf{all}}=\sum_{i=0}^{n-1}\mathbf{v}_i$ is the estimate using all prescribed reports, then dropout changes it to $\widehat{F}_{\mathsf{all}}-\sum_{i\in D}\mathbf{v}_i$.
We study the norm of this change due to dropout; the total error also includes the estimation error already present in $\widehat{F}_{\mathsf{all}}$.
For HST, the committed inputs and corrupted identities are fixed before sampling the public randomness, which is also included in the expectation.
We abbreviate selective dropout as SD and output manipulation as OM.

\begin{Thm}[Selective dropout and output manipulation]
	\label{thm:dropout-comparison}
	Let $0<\epsilon\leq1$, and assume reports are randomized independently across clients.
	For PI-RAPPOR, assume $\alpha_0q\in\mathbb{Z}$.
	For each mechanism and attack below, the worst-case expected $\ell_1$ count error is $\Theta\left(\frac{k\sqrt{n}}{\epsilon}+T\right)$, where HR denotes Hadamard Response and $T$ is:
	\begin{center}
	\setlength{\tabcolsep}{4pt}
	\begin{tabular}{lccc}
		\toprule
		Attack & (PI-)RAPPOR & HR & HST \\
		\midrule
		SD & $\frac{m\sqrt{k}}{\epsilon}$ & $\frac{m\sqrt{k}}{\epsilon}$ & $\frac{m\sqrt{k}}{\epsilon}$ \\
		OM & $\frac{mk}{\epsilon}$ & $\frac{mk}{\epsilon}$ & $\frac{m\sqrt{k}}{\epsilon}$ \\
		\bottomrule
	\end{tabular}
	\end{center}

	For (PI-)RAPPOR, the decoded contribution is $\mathbf{v}_i=\mathbf{y}_i$ as defined in Section~\ref{sec:prelim-rappor}.
	For Hadamard Response, $\mathbf{v}_i=\frac{e^\epsilon+1}{e^\epsilon-1} H_K[1:k,y_i]$, where $H_K[1:k,y_i]$ denotes the entries in rows $1,\ldots,k$ of column $y_i$.
	For HST, the decoded contribution is $\mathbf{v}_i=\mathbf{y}_i$ as defined in Section~\ref{sec:prelim-hst}.

	The following attacks achieve the attack-dependent rates in the table:
	\begin{itemize}[leftmargin=*]
		\item \textbf{Selective dropout:} Sample $s$ uniformly from $\{-1,1\}^k$ and drop each corrupted report for which $\langle s,\mathbf{v}_i\rangle\geq0$.
		This attack takes $O\left(mk\right)$ arithmetic operations, including field operations for PI-RAPPOR, and changes the estimate by $\Omega\left(\frac{m\sqrt{k}}{\epsilon}\right)$ in expected $\ell_1$ norm.
		\item \textbf{Output manipulation:} For (PI-)RAPPOR and Hadamard Response, have all corrupted clients submit the same report.
		For HST, flip the report bits selected by the selective-dropout attack.
	\end{itemize}
\end{Thm}

The common $\frac{k\sqrt{n}}{\epsilon}$ term results from the error when all clients submit correctly generated reports.
The output-manipulation comparison builds on prior analyses~\cite{CheuSU21, CaoJG21}; we analyze the selective dropout attack described above for these mechanisms.
For HST, the public vectors constrain forged reports to client-specific directions, so preventing output manipulation improves the worst-case rate by at most a constant factor.

\begin{proof}[Proof sketch]
	For (PI-)RAPPOR, covariance bounds and symmetrization control adaptive selection; for Hadamard Response, orthogonality bounds it through report collisions.
	For HST, flipping a bit changes its contribution by twice the amount of dropping it, and a random-matrix bound controls the error.
	Moment bounds and the selective dropout attack above give matching lower bounds.
	Appendix~\ref{app:dropout-comparison} gives the proofs and bounds on the change to the estimator before combining it with the error without an attack.
\end{proof}

\subsection{Comparison with Output Manipulation}

For RAPPOR, PI-RAPPOR, and Hadamard Response, comparing the selective dropout term $\frac{m\sqrt{k}}{\epsilon}$ with the error without an attack, $\frac{k\sqrt{n}}{\epsilon}$, gives the threshold $m=O\left(\sqrt{nk}\right)$.
The output-manipulation contribution $\Theta\left(\frac{mk}{\epsilon}\right)$ instead gives $m=O\left(\sqrt{n}\right)$.
These comparisons yield Corollary~\ref{cor:dropout-threshold}.

\begin{Cor}[Comparison with output manipulation]
	\label{cor:dropout-threshold}
	For RAPPOR, PI-RAPPOR, and Hadamard Response, selective dropout introduces no additional asymptotic error when
	\begin{equation}
		m=O\left(\sqrt{nk}\right).
	\end{equation}
	In contrast, output manipulation introduces no additional asymptotic error only when $m=O\left(\sqrt{n}\right)$.
\end{Cor}

The mechanisms in Corollary~\ref{cor:dropout-threshold} tolerate a $\sqrt{k}$ larger corruption budget under selective dropout than under output manipulation when $k\leq n$.
For $k>n$, the selective-dropout budget saturates at $n$, giving a $\sqrt{n}$ rather than $\sqrt{k}$ budget gain.
For HST, both attacks are absorbed by the error without an attack at the same threshold $m=O\left(\sqrt{nk}\right)$.
Thus, the benefit of verification depends on the mechanism: (PI-)RAPPOR and Hadamard Response exhibit an asymptotic gap between the two attacks, whereas the worst-case error rate is unchanged for HST.
The experiments in Section~\ref{sec:exp}, Figure~\ref{fig:main}, illustrate this distinction: for PI-RAPPOR and Hadamard Response, the gap grows with the domain, whereas the errors under output manipulation and selective dropout remain within a constant factor for HST.

\section{Post-Hoc Consistency Checks}\label{sec:cons}

The preceding sections improve robustness by requiring clients to prove correct randomization.
When changing the client protocol is not feasible, the server must instead work with reports that may have been manipulated.
We develop two \emph{consistency} checks for RAPPOR and Hadamard Response that mitigate output manipulation using statistical properties of honest reports.
Unlike verification, these checks do not certify each report's computation; they improve robustness through server-side processing alone.
They require no client-side changes or cryptography and can be applied \emph{post-hoc}, at any time after collection, including to reports from existing deployments.

As in Section~\ref{sec:setup}, $n$ is the number of clients and $m$ bounds the number of corrupted clients.
For the RAPPOR check, we assume that malicious reports are fixed independently of the honest clients' random reports.
The checks use statistical properties that hold with high probability for honest reports.
The RAPPOR check removes reports with unusually large pairwise correlations; the Hadamard Response check clips unusually large counts in the response histogram.
The guarantees below quantify the remaining error under output manipulation.
As in Equation~\eqref{eq:dropout-error}, error is measured against the count histogram of all clients' inputs, including those of corrupted clients.
The concentration arguments use Hoeffding's and Bernstein's inequalities~\cite{Vershynin18}.

\subsection{Consistency Check for RAPPOR}
\label{sec:cons-rappor}
Let $\mathbf{y}_0, \ldots, \mathbf{y}_{n-1} \in \{a,b\}^k$ be the decoded RAPPOR reports from Equation~\eqref{eq:rappor-report}, with $a = -\frac{1}{e^{\frac{\epsilon}{2}}-1}$ and $b = \frac{e^{\frac{\epsilon}{2}}}{e^{\frac{\epsilon}{2}}-1}$.
For each pair of decoded reports with $i\neq j$, the server checks whether
\begin{equation}
    |\langle \mathbf{y}_i, \mathbf{y}_j \rangle| \leq \tau,
    \qquad \tau = b^2 + b(b-a)\sqrt{2k\log n}.
    \label{eq:rappor-check-threshold}
\end{equation}
The following lemma bounds the probability that an honest report fails this check against a fixed vector.
\begin{Lem}\label{lem:rappor-check}
    Suppose $\mathbf{y} \in \{a,b\}^k$ is obtained from an honest execution of RAPPOR, and let $\mathbf{z} \in \{a,b\}^k$ be a fixed vector.
    Then, $\prob{|\langle \mathbf{y}, \mathbf{z} \rangle| \geq \tau} \leq \frac{2}{n^4}$.
\end{Lem}
\begin{proof}[Proof sketch of Lemma~\ref{lem:rappor-check}]
    All but the input coordinate give independent mean-zero summands.
    Hoeffding's inequality controls their sum, while the input coordinate contributes at most $b^2$.
    See Appendix~\ref{app:rappor-check-proof} for the full proof.
\end{proof}
For malicious reports fixed independently of the honest clients' random reports, a union bound shows that no pair involving an honest report fails the check except with probability at most $\frac{2}{n^2}$.
The server therefore retains exactly the decoded reports whose absolute inner product with every other original decoded report is at most $\tau$, and sums them to obtain a histogram estimate.

A full check takes $O\left(n^2k\right)$ time.
If an upper bound $1\leq m\leq n$ on the number of corrupted clients is known, the server can reduce the number of pairs it checks as follows.
After all reports are fixed, the server samples $s=\min\left\{n,\ceil{\frac{n\sqrt{k}}{m}}\right\}$ reports uniformly without replacement, independently of the reports.
It compares each of these $s$ sampled reports with every other original report and discards both reports in every detected violating pair.
This takes $O\left(n^2k\min\left\{1,\frac{\sqrt{k}}{m}\right\}\right)$ time and preserves the full check's expected-error bound, where the expectation also includes the server's sampling.

\begin{Thm}\label{thm:rappor-consistency}
    For $0<\epsilon\leq1$, the full and sampled RAPPOR consistency checks have $O\left(\frac{k \sqrt{n}}{\epsilon} + \frac{m k^{\frac{3}{4}} \log(n)^{\frac{1}{4}}}{\epsilon}\right)$ expected $\ell_1$ error under output manipulation, with malicious reports fixed independently of the honest reports.
\end{Thm}
\begin{proof}[Proof sketch of Theorem~\ref{thm:rappor-consistency}]
    With high probability, no pair involving an honest report violates the check, so honest reports survive.
    In the full check, the retained malicious reports have bounded pairwise correlations, which control the $\ell_2$ norm of their sum.
    Cauchy--Schwarz converts this bound to $\ell_1$ error, to which we add honest sampling error.
    For the sampled check, fix reports for which no pair involving an honest report violates the check.
    The expected number of violating pairs whose reports both survive, over the server's sample of $s$ reports, is at most $\frac{mn}{2es}$.
    This bound suffices to obtain the same expected-error rate.
    Appendix~\ref{app:rappor-consistency} gives the full proof.
\end{proof}

\subsection{Consistency Check for Hadamard Response}

Let $y=(y_0,\ldots,y_{n-1})\in\{0,\ldots,K-1\}^n$ denote the Hadamard Response reports, and let $\mathbf{c}(y)\in\mathbb{Z}_{\geq0}^K$ be their report histogram, where $\mathbf{c}(y)_j=\sum_{i=0}^{n-1}\mathbbm{1}[y_i=j]$.
This histogram counts report symbols, not the clients' input values.
The consistency check limits each count using the threshold
\begin{equation}
    T_{\mathsf{HR}}=2n\left(\frac{2e^\epsilon}{(e^\epsilon+1)K}\right) + 5 \log(nK).
    \label{eq:had-check-threshold}
\end{equation}
This limits the contribution of any one report symbol to the estimator.
Concentration of independent report indicators gives the following guarantee.

\begin{Lem}\label{lem:had-check}
    Suppose $y_0,\ldots,y_{n-1}$ are obtained from an honest execution of Hadamard Response.
    Then, $\prob{\forall j\in\{0,\ldots,K-1\},\;\mathbf{c}(y)_j \leq T_{\mathsf{HR}}} \geq 1-\frac{1}{n}$.
\end{Lem}
\begin{proof}[Proof sketch of Lemma~\ref{lem:had-check}]
    Each count is a sum of independent Bernoulli indicators and has mean at most $\frac{2ne^\epsilon}{(e^\epsilon+1)K}$.
    Bernstein's inequality and a union bound over the $K$ counts give the claim.
    See Appendix~\ref{app:hadamard-check-proof} for the full proof.
\end{proof}

The server clips each count to $\tilde{\mathbf{c}}(y)_j=\min\left\{\mathbf{c}(y)_j,T_{\mathsf{HR}}\right\}$, then applies the standard Hadamard Response decoder to the clipped histogram $\tilde{\mathbf{c}}(y)$.
The resulting estimator has the following expected-error bound under output manipulation:

\begin{Thm}\label{thm:hadamard-consistency}
    For $0<\epsilon\leq1$, Hadamard Response with the above consistency check has $O\left(\frac{k\sqrt{n}}{\epsilon} + \frac{k\sqrt{m \log(nk)}}{\epsilon} + \frac{\sqrt{kmn}}{\epsilon}\right)$ expected $\ell_1$ error under output manipulation.
\end{Thm}
\begin{proof}[Proof sketch of Theorem~\ref{thm:hadamard-consistency}]
    Let $\mathbf{c}_{H}$ be the histogram of honest reports alone.
    With high probability, every entry of $\mathbf{c}_{H}$ is at most $T_{\mathsf{HR}}$.
    On this event, clipping leaves the honest counts intact, and the residual $\tilde{\mathbf{c}}(y)-\mathbf{c}_{H}$ has $\ell_1$ norm at most $m$ and $\ell_\infty$ norm at most $T_{\mathsf{HR}}$.
    Hadamard orthogonality bounds its decoded $\ell_2$ norm, and Cauchy--Schwarz bounds its $\ell_1$ contribution.
    Adding honest sampling error gives the result.
    Appendix~\ref{app:hadamard-consistency} contains the full proof.
\end{proof}

The server-side consistency check adds $O\left(n+k\right)$ work to the $O\left(n+k\log k\right)$ runtime of standard Hadamard Response.
The check also generalizes to protocols based on other design matrices, such as Projective Geometry Response~\cite{FeldmanNNT22}.

\subsection{Discussions and Limitations}
\label{sec:discussion}

Server-side consistency checks offer an alternative when changing client implementations is impractical.
They require no client-side cryptography and can be applied to reports already collected.
Their error guarantees are weaker than those under selective dropout with verification: RAPPOR's check has an $O\left(\frac{mk^{\frac{3}{4}}(\log n)^{\frac{1}{4}}}{\epsilon}\right)$ attack term, and Hadamard Response's check has an $O\left(\frac{\sqrt{kmn}+k\sqrt{m\log(nk)}}{\epsilon}\right)$ term, compared with $\Theta\left(\frac{m\sqrt{k}}{\epsilon}\right)$ under selective dropout.

The full RAPPOR check requires $O\left(n^2k\right)$ work; given a corruption budget $m\geq1$, the sampled variant reduces this to $O\left(n^2k\min\{1,\frac{\sqrt{k}}{m}\}\right)$ with the same expected-error bound.
Even this cost can be substantial for large populations or domains.
Constant-factor distance approximations need not preserve the much smaller inner-product tolerance used by the RAPPOR check.
The RAPPOR analysis also relies on independent report coordinates, so its guarantee does not directly extend to PI-RAPPOR.
By contrast, the Hadamard check adds only $O\left(n+k\right)$ work to aggregation.
Further reducing the server-side cost and developing better checks for both mechanisms, as well as comparable guarantees for PI-RAPPOR and other LDP mechanisms, remain open questions.
\section{Experiments}\label{sec:exp}

We show that our verifiable PI-RAPPOR and Hadamard Response implementations from Section~\ref{sec:protocol} are lightweight and efficient.
We also evaluate the effects of selective dropout and output manipulation on the accuracy of (PI-)RAPPOR, Hadamard Response, and HST across input distributions, as well as the accuracy obtained using the server-side consistency checks for RAPPOR and Hadamard Response from Section~\ref{sec:cons}.

\subsection{Experimental Setup}

\noindent\textbf{Implementation and platform.}
We implement the mechanisms and verification circuits in Rust, using arkworks 0.4~\cite{Arkworks22}, Groth16~\cite{Groth16} over BLS12-381, and Poseidon-based commitments~\cite{GrassiKRRS21}.
The per-client benchmark requests 32 CPUs and 64\,GiB of memory on an Ubuntu 22.04 machine with two Intel Xeon Platinum 8358 processors at 2.60\,GHz (64 physical cores).
We evaluate $k\in\{3,15,63,255,1023,4095\}$, giving Hadamard output alphabets of size $K=k+1$.
Other domain sizes are supported by padding to $K=2^{\lceil\log_2(k+1)\rceil}$ and using the first $k$ nonzero Hadamard rows.
Both circuits use $\gamma=12$ bits for random threshold comparisons, and the accuracy estimators correct for the resulting rounded probabilities.

\noindent\textbf{Input distributions.}
We use uniform, truncated geometric with weights proportional to $0.8\cdot0.2^j$, and truncated Zipf with weights proportional to $(j+1)^{-1.1}$, normalized over $j\in\{0,\ldots,k-1\}$.
For the Kosarak clickstream dataset from the SPMF repository\footnote{\texttt{https://www.philippe-fournier-viger.com/}\allowbreak\texttt{spmf/}}~\cite{Fournier-VigerG14}, we count page occurrences, retain the $k-1$ most frequent pages, and pool the rest into an \emph{Other} category.
Client inputs are sampled independently with replacement from these distributions; within each trial, all mechanisms use the same input seed, independently of mechanism randomness.

\noindent\textbf{Error metric.}
We measure the $\ell_1$ error $\|\widehat F-F(\mathbf{x})\|_1$ of the estimated count histogram against all $n$ inputs, including those of dropped clients.
We sum the decoded reports that are submitted without normalizing by their number or projecting onto a probability simplex.

\subsection{Per-client Latency}

\noindent\textbf{Measurement procedure.}
We benchmark the verifiable protocols for PI-RAPPOR and Hadamard Response from Section~\ref{sec:protocol} over the six domain sizes and $\epsilon\in\{\frac{1}{8},\frac{1}{4},\frac{1}{2},1,2,4,8\}$, repeating each of the 84 configurations 100 times.
Each trial performs a fresh Groth16 setup and verification-key preprocessing for both proofs, followed by one report-generation and verification execution.
Setup is measured separately and excluded from online latency.
Prover time includes commitment computation, native report computation, and generation of the input-commitment and evaluation proofs; verifier time includes verification of both proofs but excludes aggregation.

\noindent\textbf{Communication model.}
We combine measured computation with modeled transmission for the three messages: client commitments and their knowledge proof, server randomness shares, and the report and its evaluation proof.
The symmetric network profiles are WAN (20\,ms RTT, 20\,Mbit/s), mobile (50\,ms, 10\,Mbit/s), and constrained mobile (100\,ms, 1\,Mbit/s).
Messages use canonical compressed serialization, with server bit shares represented as field elements; the model excludes connection establishment and transport overhead.

\begin{figure}
    \centering
    \includegraphics[width=\linewidth]{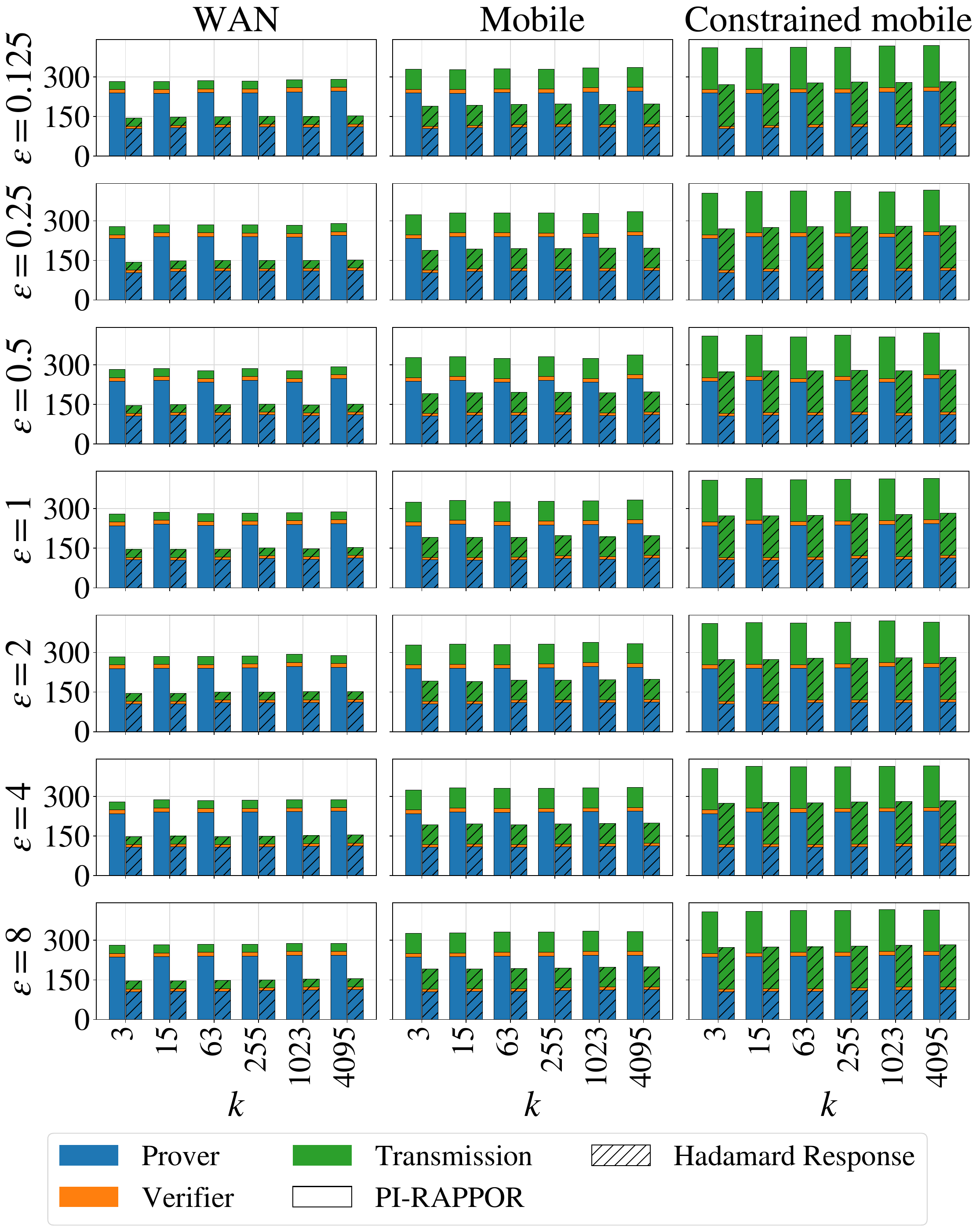}
    \caption{Mean modeled per-client online latency (ms) over 100 trials,
    including both the input-commitment and evaluation proofs.
    Setup and aggregation are excluded.}
    \label{fig:per-client-latency}
\end{figure}

\noindent\textbf{Results.}
Figure~\ref{fig:per-client-latency} shows modest variation across $k$ and $\epsilon$.
Across configurations, mean prover and verifier times are 233.2--247.5\,ms and 13.8--15.5\,ms for PI-RAPPOR, and 104.2--113.5\,ms and 9.1--10.6\,ms for Hadamard Response, respectively.
The two proof circuits together contain 6,941--7,001 constraints for PI-RAPPOR and 2,374--2,544 for Hadamard Response, explaining the small change in cryptographic costs from $k=3$ to $k=4095$.

Total communication is 992 bytes for PI-RAPPOR and 928--1,248 bytes for Hadamard Response.
Mean online latency on WAN, mobile, and constrained mobile is 277.8--293.1\,ms, 323.2--338.5\,ms, and 405.3--420.7\,ms for PI-RAPPOR; the respective Hadamard Response ranges are 143.7--154.1\,ms, 189.0--199.6\,ms, and 270.7--283.6\,ms.
On slower networks, the 1.5 round trips account for much of the increased latency because messages remain small.
Setup takes 268.6--284.2\,ms for PI-RAPPOR and 139.7--148.7\,ms for Hadamard Response.

\noindent\textbf{Comparison with V$\epsilon$rity.}
V$\epsilon$rity~\cite[Section~9.2]{BellClarkGLRC25} reports 1,423.50\,s of online client computation and 1,472.77\,s of server computation for binary randomized response, approximately 23.7 and 24.5 minutes, respectively, per client.
Our plaintext verification circuits instead handle categorical inputs with up to 4,095 values at millisecond-scale prover and verifier costs.

At the server, one million separate invocations at V$\epsilon$rity's reported server time would take about 50 years of serial execution; extending an HE-based design to our more complex categorical randomizers could increase this to the scale of a century.
Projecting our measured verification times for both proofs to one million reports gives approximately 2.5--4.3 hours, illustrating the computational advantage while retaining the quantified robustness benefit against selective dropout.

\subsection{Robustness Against Different Attacks}
For the class-size sweep, we fix $\epsilon=1$ and vary $k$ over the domain sizes above.
Appendix~\ref{app:rappor-results} reports the RAPPOR results, including the full and sampled consistency checks.
For PI-RAPPOR, Hadamard Response, and HST, we vary $n\in\{10^4,10^5,10^6\}$.
For each configuration and input distribution, we run ten independent randomized trials comparing a no-attack setting (Clean), selective dropout (SD), and output manipulation (OM) using the same inputs and valid randomized reports; we report the mean $\ell_1$ error.
In both attacks, the first $m=0.1n$ clients are corrupted before reports or HST public vectors are sampled.
We use this corruption rate to make the attacks visible and apply the strategies in Theorem~\ref{thm:dropout-comparison}.

The Clean setting measures estimation error when all clients submit correctly randomized reports.
The SD setting evaluates the attack permitted by our verification framework (Section~\ref{sec:protocol}), while the OM setting evaluates unrestricted output manipulation without verification.
For Hadamard Response, we also evaluate the histogram-clipping consistency check from Section~\ref{sec:cons}, labeled OM-CC.
\begin{figure}[!t]
\centering

\subfloat[PI-RAPPOR\label{fig:pi-rappor-l1}]{%
    \includegraphics[width=\linewidth]{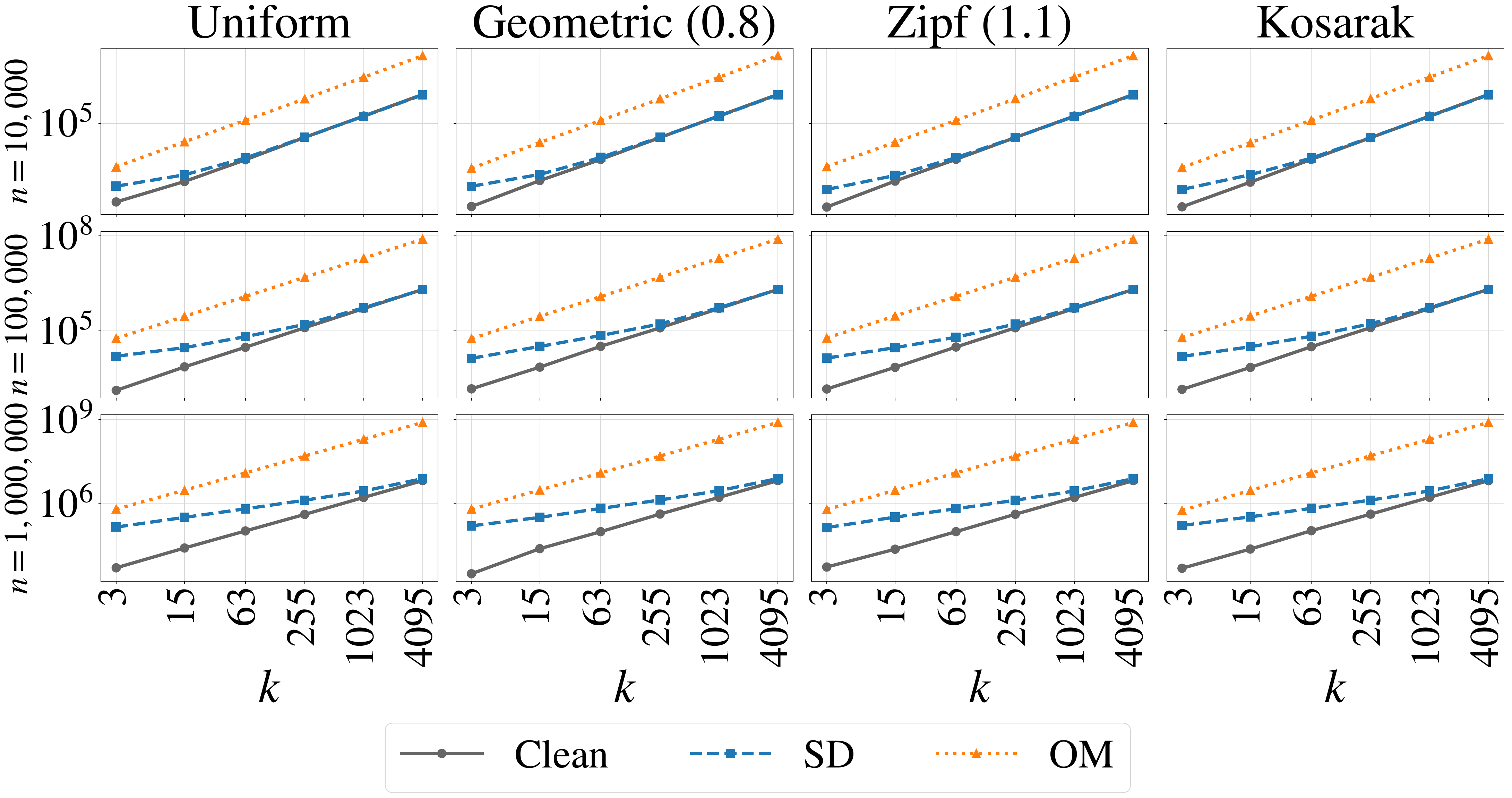}%
}\par
\subfloat[Hadamard Response\label{fig:hadamard-l1}]{%
    \includegraphics[width=\linewidth]{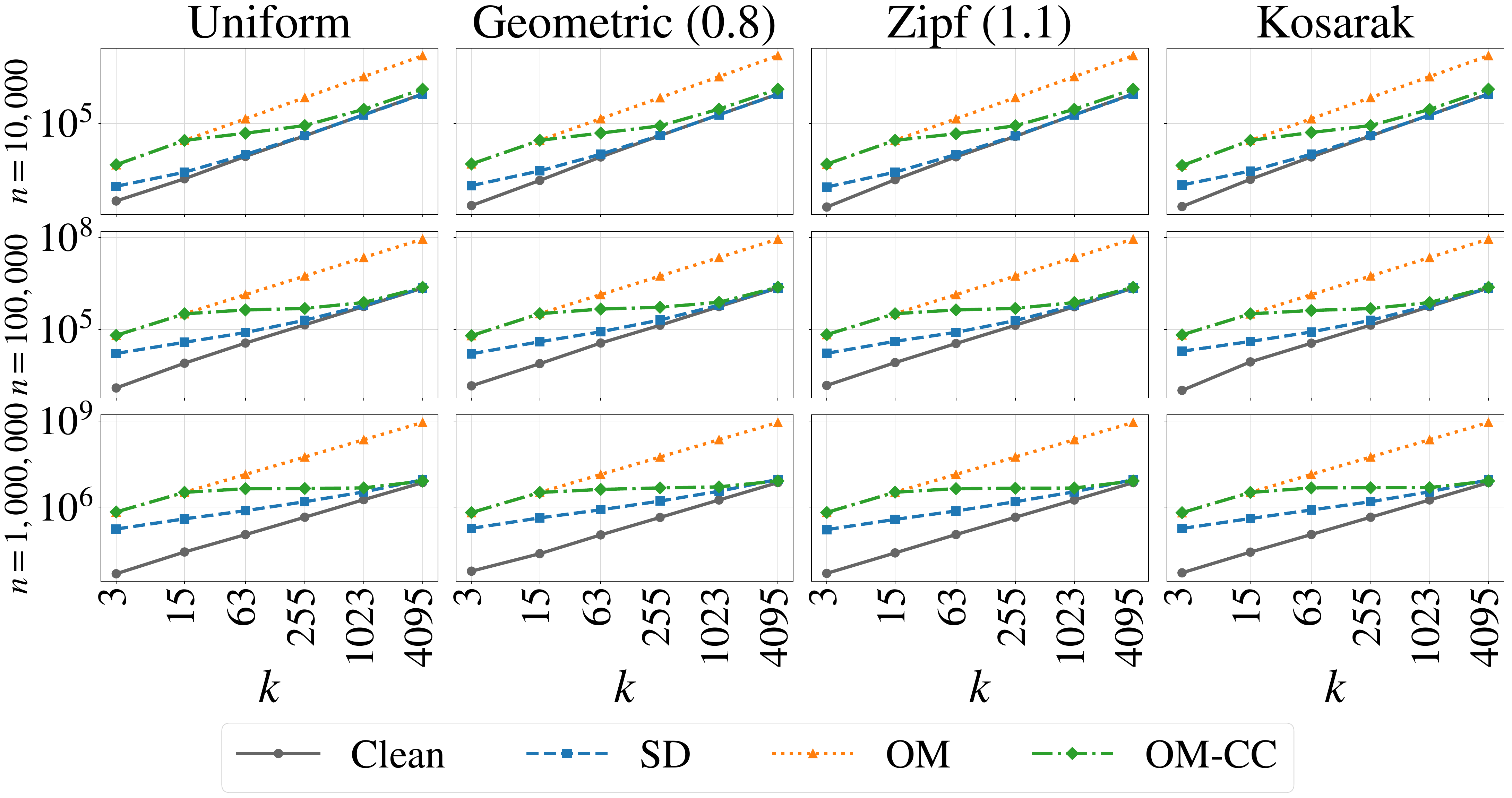}%
}\par
\subfloat[HST\label{fig:hst-l1}]{%
    \includegraphics[width=\linewidth]{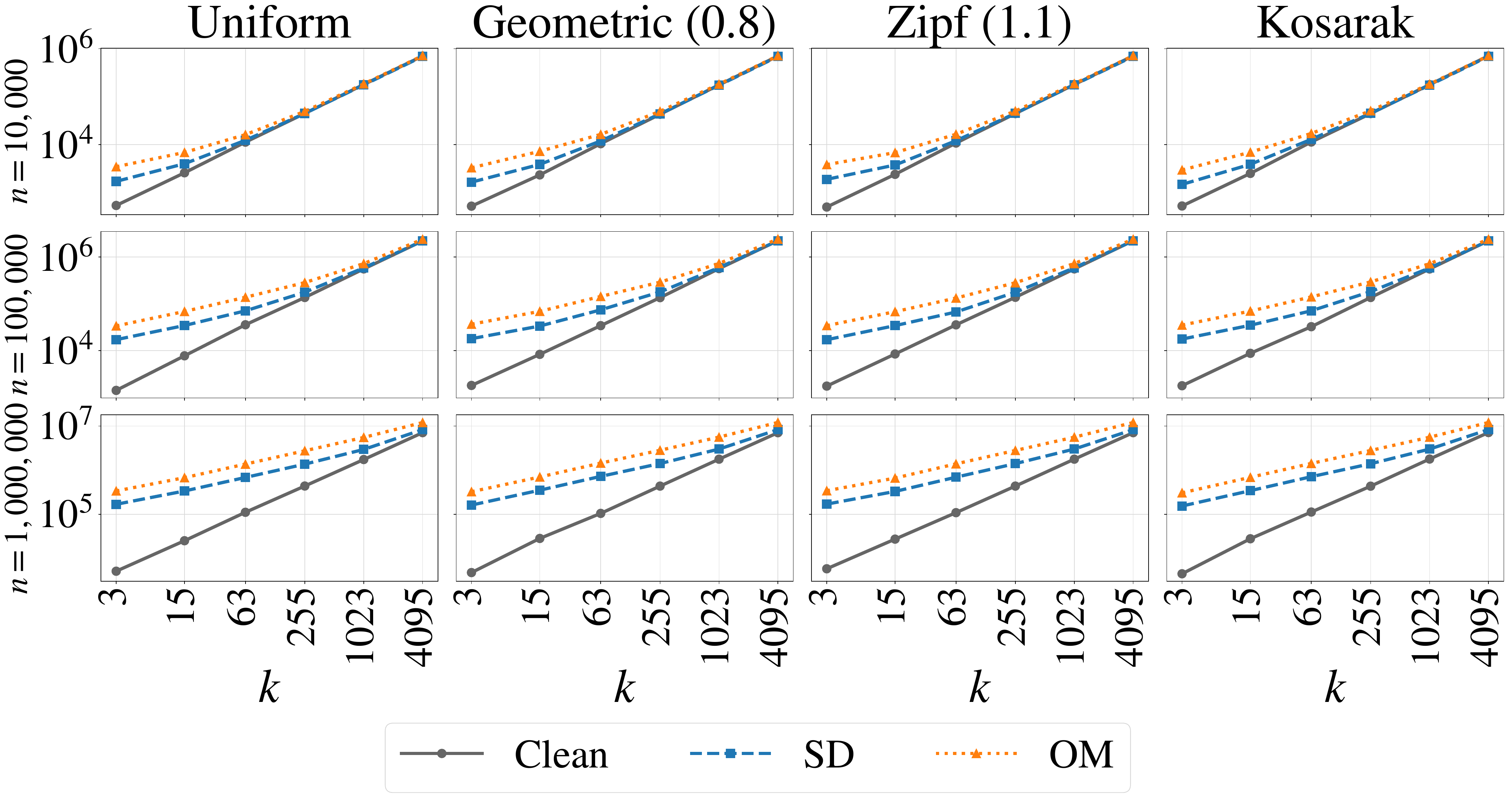}%
}\par

\caption{Mean $\ell_1$ error of the estimated count histogram over ten trials at
$\epsilon=1$. In attack trials, the adversary controls 10\% of clients.
OM and SD denote output manipulation and selective dropout; OM-CC adds the histogram-clipping consistency check for Hadamard Response.} \label{fig:main}
\end{figure}

\noindent\textbf{Results.}
Figure~\ref{fig:main} shows the mechanism-dependent distinction predicted by our analysis.
PI-RAPPOR and Hadamard Response show a much larger separation than HST as the domain grows.
At $n=10^6$ and $k=4095$, across all four distributions, PI-RAPPOR's selective-dropout error is only 1.18--1.19 times the Clean error, whereas output manipulation produces 121.5--122.3 times the Clean error.
For Hadamard Response, selective dropout and output manipulation produce 1.26--1.29 and 125--126 times the Clean error, respectively.
Thus, by restricting the adversary to selective dropout rather than output manipulation, our proposed verification framework reduces the error by about two orders of magnitude in these settings, bringing it close to the Clean error.

These trends follow Corollary~\ref{cor:dropout-threshold}: for (PI-)RAPPOR and Hadamard Response, selective dropout preserves the error rate without an attack when the corrupted fraction is $O\left(\sqrt{\frac{k}{n}}\right)$, whereas output manipulation requires $O\left(\frac{1}{\sqrt n}\right)$.
At a fixed 10\% corruption rate, increasing $k$ brings selective-dropout error closer to the Clean error while output manipulation remains far above it.
Increasing $n$ at fixed $k$ lowers both tolerable fractions, making the attacks more pronounced relative to the baseline.

For Hadamard Response, the histogram-clipping consistency check substantially reduces the error from the repeated-report attack: at $n=10^6$ and $k=4095$, mean error is 1.13--1.15 times the Clean error across the four distributions.
At $k=3$, the clipping threshold is too large to suppress the attack at the evaluated corruption rate, so the checked and unchecked errors coincide.
As the output alphabet grows, the threshold decreases relative to the number of repeated malicious reports, making clipping more effective.

For HST, the errors under selective dropout and output manipulation remain within a constant factor: with uniform inputs, $n=10^6$, and $k=4095$, their mean errors are 1.17 and 1.70 times the Clean error, respectively.

Appendix~\ref{app:corrupted-fraction} presents results for varying the corrupted fraction.
As the domain size increases, selective-dropout error remains close to the Clean error over a wider range of corrupted fractions, consistent with Corollary~\ref{cor:dropout-threshold}.

\noindent\textbf{Comparison with HST.}
At $\epsilon=1$ and a 10\% corruption rate, HST's mean output-manipulation error is 1.13--2.71 times the selective-dropout error of verified PI-RAPPOR and 1.03--2.58 times that of verified Hadamard Response across our evaluated configurations.
Appendix~\ref{app:hst-comparison} reports the full ratios.
Our current verified implementations transmit more bytes than bit-packed HST over the evaluated domain sizes, reflecting a tradeoff between accuracy and communication.

\section{Conclusion}
\label{sec:conclusion}

Correct execution does not prevent selective withholding of valid outputs.
For (PI-)RAPPOR and Hadamard Response, restricting the adversary to selective dropout rather than output manipulation reduces the attack-dependent error by a factor of $\Theta\left(\sqrt{k}\right)$; HST exhibits no such asymptotic separation.

Thus, verification can improve robustness without eliminating selective dropout, with mechanism-dependent benefits. 
Our lightweight framework uses commitments and proofs of plaintext evaluation to prevent output manipulation while allowing selective dropout.
When client changes are impractical, server-side consistency checks offer a post-hoc alternative with weaker error guarantees.

Future work includes extending our selective-dropout analysis to other LDP mechanisms and developing stronger and more scalable consistency checks. 
More broadly, our findings highlight the need for robustness analyses of verifiable DP to account for selective dropout.

\bibliographystyle{ACM-Reference-Format}
\bibliography{reference}

\appendix
\section{Additional Experiments}
\subsection{Additional Results for RAPPOR}
\label{app:rappor-results}

We evaluate RAPPOR using the domain sizes, input distributions, error metric, and attack strategies described in Section~\ref{sec:exp}.
We fix $\epsilon=1$ and use $n=10^4$ clients due to computational resource constraints, with $m=0.1n$ corrupted clients in attack trials.
Each configuration is evaluated over ten independent randomized trials.
We compare a no-attack setting (Clean), selective dropout (SD), and output manipulation (OM), and evaluate the full and sampled consistency checks from Section~\ref{sec:cons}, labeled OM-CC and OM-CC-S, respectively.

\begin{figure}[!t]
    \centering
    \includegraphics[width=\linewidth]{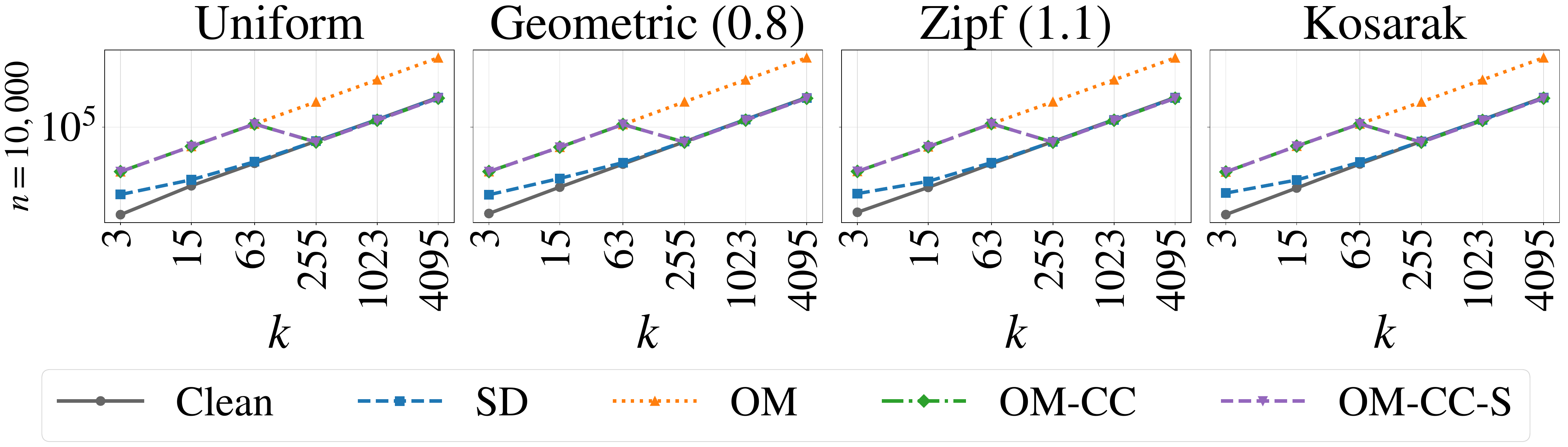}
    \caption{Mean $\ell_1$ error of the estimated count histogram for RAPPOR over ten trials at $n=10^4$ and $\epsilon=1$.
    In attack trials, the adversary controls 10\% of clients.
    OM and SD denote output manipulation and selective dropout; OM-CC and OM-CC-S add the full and sampled consistency checks, respectively.}
    \label{fig:rappor-l1}
\end{figure}

At $n=10^4$ and $k=4095$, Figure~\ref{fig:rappor-l1} shows that selective-dropout error is close to the Clean error across the four distributions, while output manipulation produces about 12.1 times that error.

Both the full and sampled consistency checks yield mean errors approximately 5\% below the Clean error at $k=4095$.
Filtering removes noisy contributions as well as malicious reports; at these parameters, the reduction in sampling noise can outweigh the missing-input bias.
The sharp improvement between $k=63$ and $k=255$ is consistent with threshold-based filtering of the repeated-report attack.
At $n=10^4$ and $\epsilon=1$, the expected squared norm of the repeated report increases from approximately 248 to 1,000, while the correlation threshold increases from approximately 360 to 718.
This makes repeated reports much more likely to violate the check.

\subsection{Varying the Corrupted Fraction}
\label{app:corrupted-fraction}

\begin{figure}[!t]
    \centering
    \includegraphics[width=\linewidth]
        {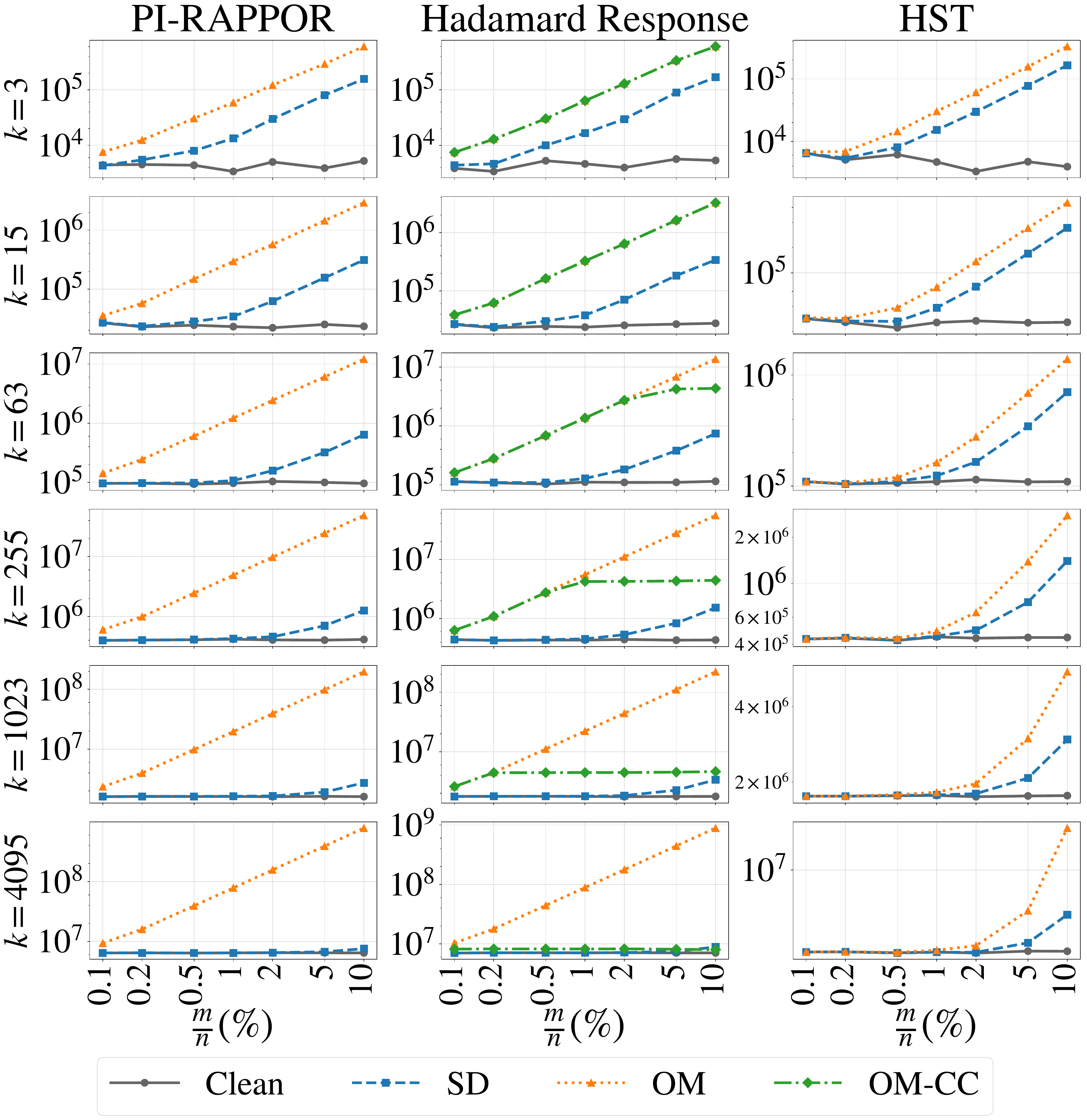}
    \caption{Mean $\ell_1$ error of the estimated count
    histogram over ten trials with uniform inputs,
    $n=10^6$, and $\epsilon=1$.}
    \label{fig:m-sweep}
\end{figure}

In Figure~\ref{fig:m-sweep}, the selective-dropout curve separates from the Clean curve at larger corrupted fractions as $k$ increases, consistent with the threshold scale $m=\Theta\left(\sqrt{nk}\right)$ from Corollary~\ref{cor:dropout-threshold}.
For fixed $n$, this corresponds to $\frac{m}{n}=\Theta\left(\sqrt{\frac{k}{n}}\right)$.
For example, at $\frac{m}{n}=1\%$ and $k=3$, selective-dropout error is about 4.0 and 3.6 times the error without an attack for PI-RAPPOR and Hadamard Response, respectively; at $k=4095$, it remains close to that error for both.
The HST curves show the same shift with $k$, while its output-manipulation error remains closer to its selective-dropout error.

For Hadamard Response, the error with the consistency check matches the unchecked output-manipulation error while the total count in the attacked report bin remains below the clipping threshold.
For fixed $n$ and $k$, this threshold does not depend on $m$.
Once the count exceeds the threshold, additional copies are clipped, limiting the growth of the attack's contribution.
As $k$ increases, the threshold decreases, so clipping becomes effective at smaller corrupted fractions.
At $k=3$, the check does not reduce the attack's error over the evaluated range; at $k=4095$, checked error remains approximately 1.13--1.17 times the Clean error throughout the sweep.

\subsection{Comparison with HST}
\label{app:hst-comparison}

Table~\ref{tab:hst-empirical-ratios} compares HST under output manipulation with verified PI-RAPPOR and Hadamard Response under selective dropout, using the class-sweep experiments in Section~\ref{sec:exp}.
Each ratio is computed from the corresponding ten-trial mean $\ell_1$ errors of the estimated count histograms at $\epsilon=1$ and $m=0.1n$.
For each $n$ and $k$, the table gives the minimum and maximum across the four input distributions.
Within each cell, the upper line gives the ratio for PI-RAPPOR and the lower line gives the ratio for Hadamard Response.
These results describe the evaluated attacks and parameter settings; they do not establish an asymptotic separation between HST and the verified mechanisms.

\begin{table}[!t]
    \centering
    \setlength{\tabcolsep}{3pt}
    \newcommand{\hstpair}[2]{%
        \begin{tabular}[c]{@{}c@{}}
            #1\\
            #2
        \end{tabular}%
    }
    \caption{Ratios of mean $\ell_1$ errors of the estimated count histograms
    at $\epsilon=1$ with $m=0.1n$ corrupted clients.
    In each cell, the upper line divides HST's
    output-manipulation error by the selective-dropout
    error of verified PI-RAPPOR; the lower line uses
    verified Hadamard Response (HR).
    Each range gives the minimum and maximum across Uniform,
    Geometric(0.8), Zipf(1.1), and Kosarak inputs,
    computed from ten-trial means for each distribution;
    the ranges are not confidence intervals.
    Inputs are matched across mechanisms.
    A ratio above one indicates lower error for the
    verified mechanism. Errors include sampling error.}
    \label{tab:hst-empirical-ratios}
    \begin{tabular}{cccc}
        \hline
        \hstpair{PI-RAPPOR}{HR}
        & $n=10^4$ & $n=10^5$ & $n=10^6$ \\
        \hline
        $k=3$
        & \hstpair{1.895--2.711}{1.742--2.581}
        & \hstpair{2.109--2.660}{1.794--2.268}
        & \hstpair{1.919--2.553}{1.722--2.104} \\[3pt]
        $k=15$
        & \hstpair{1.896--1.960}{1.625--1.738}
        & \hstpair{2.124--2.305}{1.654--1.800}
        & \hstpair{2.081--2.249}{1.678--1.796} \\[3pt]
        $k=63$
        & \hstpair{1.467--1.607}{1.253--1.312}
        & \hstpair{2.001--2.094}{1.651--1.743}
        & \hstpair{2.166--2.221}{1.806--1.911} \\[3pt]
        $k=255$
        & \hstpair{1.207--1.259}{1.084--1.131}
        & \hstpair{1.717--1.767}{1.424--1.483}
        & \hstpair{2.148--2.187}{1.760--1.829} \\[3pt]
        $k=1023$
        & \hstpair{1.139--1.190}{1.027--1.062}
        & \hstpair{1.333--1.349}{1.172--1.195}
        & \hstpair{2.000--2.037}{1.587--1.650} \\[3pt]
        $k=4095$
        & \hstpair{1.126--1.136}{1.035--1.042}
        & \hstpair{1.177--1.197}{1.069--1.076}
        & \hstpair{1.570--1.589}{1.315--1.358} \\[3pt]
        \hline
    \end{tabular}
\end{table}

\section{Proofs}
\label{app:proofs}

\subsection{Proof of Theorem~\ref{thm:client-predictable-bias}}
\label{app:client-predictable-bias}
\begin{proof}
	After receiving $r^{\ser}$, the malicious client runs the honest client algorithm independently
	\begin{equation}
		T=\kappa p(\kappa)
	\end{equation}
	times, each time with fresh client randomness.
	For each execution, it uses $\mathsf{ComputeReport}$ to compute the corresponding report $y$ from its local view.
	It sends the response from the first trial for which $\mathsf{ComputeReport}$ returns a report in $S$; if no such response is found, it sends the response from the first trial.
	Each trial follows the prescribed report and proof computations; the deviation is choosing which trial to submit rather than submitting the result of a single trial.

	For each server challenge satisfying the assumptions, the probability that none of the $T$ trials generates a report in $S$ is at most
	\begin{equation}
		\left(1-\frac{1}{p(\kappa)}\right)^T
		\leq e^{-\kappa}.
	\end{equation}
	By a union bound over the $T$ trials, the probability that $\mathsf{ComputeReport}$ computes any report incorrectly remains negligible.
	The selected response is generated by the honest client algorithm and its report satisfies $y=f(x,r^{\cli},r^{\ser})$ for that execution's randomness, so the associated proof remains valid rather than being forged.
	By correctness of $\Pi$, rejection of an honestly generated response has negligible probability; a union bound over the $T$ trials preserves this guarantee for the selected response.
	Accounting also for the negligible probability of a server challenge that fails the assumptions, the server accepts the selected response with report $y\in S$ except with negligible probability.
	In contrast, after averaging over the server's challenge, an honest execution generates a report in $S$ with probability at most $\frac{1}{2}+\mathsf{negl}(\kappa)$ by Equation~\eqref{eq:predictable-event}.
	The difference on the event $S$ gives the claimed total-variation-distance bound.
\end{proof}

\subsection{Proof of Theorem~\ref{thm:lightweight-protocol}}
\label{app:lightweight-protocol}
\begin{proof}
	Completeness follows directly from correctness of the commitment and completeness of the two ZKPoKs.
	For accepted-output integrity, the first proof establishes knowledge of a valid opening of $c_i$ before the server samples $r_i^{\ser}$.
	Binding fixes the committed input and client randomness share.
	The second proof then guarantees that an accepted $y_i$ satisfies Equation~\eqref{eq:evaluation-relation} for that same input and randomness share.
	Therefore, after receiving $r_i^{\ser}$, a client has only one report that can be accepted for its commitment: it may submit that report or drop out.

	For privacy, the server's view of an honest client consists of $(c_i,\pi_i^{\mathsf{in}},r_i^{\ser},y_i,
	\pi_i^{\mathsf{eval}})$. Hiding of the commitment and zero knowledge of
	the proofs ensure that the commitment and proofs leak no additional information about $(x_i,r_i^{\cli})$ beyond the public transcript $(r_i^{\ser},y_i)$.
	By the distributed-randomness requirement, $(r_i^{\ser},y_i)$ is $\epsilon$-computationally locally private with respect to $x_i$.
	Closure under PPT post-processing gives the same guarantee for the complete server view.
\end{proof}

\subsection{Proof of Theorem~\ref{thm:dropout-comparison}}
\label{app:dropout-comparison}

\noindent\textbf{Common bounds.}
Write $\mathbf{v}_i$ for a decoded report and
$E_0=\sum_{i=0}^{n-1}(\mathbf{v}_i-\mathbf{e}_{x_i})$.
For all four mechanisms and $0<\epsilon\leq1$, each centered coordinate
has variance $\Theta\left(\epsilon^{-2}\right)$ and fourth moment $O\left(\epsilon^{-4}\right)$:
for (PI-)RAPPOR this follows from the Bernoulli marginals, and for HR and
HST from $v_{i,j}\in\{-\lambda_\epsilon,\lambda_\epsilon\}$,
$\lambda_\epsilon=\frac{e^\epsilon+1}{e^\epsilon-1}$, and
$\mathbb{E}[v_{i,j}]=\mathbbm{1}[x_i=j]$.
Thus a sum $W$ of $r\geq1$ independent centered coordinates has
$\mathbb{E}[W^2]=\Theta\left(\frac{r}{\epsilon^2}\right)$ and
$\mathbb{E}[W^4]=O\left(\frac{r^2}{\epsilon^4}\right)$.
Cauchy--Schwarz and Paley--Zygmund applied to $W^2$ give
\begin{equation}
	\mathbb{E}|W|=\Theta\left(\frac{\sqrt r}{\epsilon}\right),
	\qquad
	\mathbb{E}\|E_0\|_1=\Theta\left(\frac{k\sqrt n}{\epsilon}\right).
	\label{eq:rappor-moment-estimate}
\end{equation}
Only independence across clients is used; for HST the expectation also
includes the public randomness.

Let $L=\max_{D\subseteq C}\|\sum_{i\in D}\mathbf{v}_i\|_1$.
Every realized report has $\|\mathbf{v}_i\|_2=\Theta\left(\frac{\sqrt{k}}{\epsilon}\right)$.
For an independent uniform sign vector $s\in\{-1,1\}^k$, symmetry and
Khintchine's inequality imply
\begin{equation}
	\mathbb{E}_s(\langle s,\mathbf{v}_i\rangle)_+
	=\tfrac12\mathbb{E}_s|\langle s,\mathbf{v}_i\rangle|
	=\Theta\left(\frac{\sqrt k}{\epsilon}\right).
	\label{eq:rappor-sign-attack}
\end{equation}
Dropping reports with nonnegative projection therefore gives
$\mathbb{E}L=\Omega\left(\frac{m\sqrt{k}}{\epsilon}\right)$.
Taking $D=C$ gives the other lower bound:
for each coordinate, write the sum as $\mu+W$, where $W$ is centered.
Since
$\mathbb{E}|\mu+W|\geq\max\{|\mu|,\mathbb{E}|W|-|\mu|\}
\geq\frac12\mathbb{E}|W|$,
Equation~\eqref{eq:rappor-moment-estimate} yields
$\mathbb{E}L=\Omega\left(\frac{k\sqrt m}{\epsilon}\right)$.
Together these imply
$\mathbb{E}L=\Omega\left(\frac{m\sqrt{k}+k\sqrt m}{\epsilon}\right)$.

To pass from perturbation to total error, use
$\widehat F-F(\mathbf{x})=E_0-\sum_{i\in D}\mathbf{v}_i$.
An upper bound of the same order on $\mathbb{E}L$, together with $m\leq n$,
gives $O\left(\frac{k\sqrt n+m\sqrt k}{\epsilon}\right)$ total error.
For the lower bound, full participation attains the error without an attack,
while the selective dropout attack in Theorem~\ref{thm:dropout-comparison} satisfies
\begin{equation}
	\begin{aligned}
		\mathsf{Err}_1
		&\geq\mathbb{E}[-\langle s,\widehat F-F(\mathbf{x})\rangle]\\
		&=\sum_{i\in C}\mathbb{E}(\langle s,\mathbf{v}_i\rangle)_+
		=\Omega\left(\frac{m\sqrt k}{\epsilon}\right),
	\end{aligned}
	\label{eq:sign-total-error}
\end{equation}
because $s$ is independent of the mean-zero $E_0$.
The maximum of the two lower bounds is at least half their sum.
The attack costs $O\left(mk\right)$ operations; when $m=0$, only the error without an attack remains.

\subsubsection{(PI-)RAPPOR}
\label{app:rappor-proof}
\begin{proof}
	Set $a=\alpha_0=(1+e^{\frac{\epsilon}{2}})^{-1}$ and
	$\Delta=1-2a=\Theta\left(\epsilon\right)$, so $y_j=\frac{Z_j-a}{\Delta}$.
	For RAPPOR, coordinate independence gives
	$\|\operatorname{Cov}(\mathbf{y}\mid x)\|_{\mathsf{op}}=O\left(\epsilon^{-2}\right)$.
	For PI-RAPPOR, let $U$ be uniform on $\F_q^2$.
	The bijection $(u,v)\mapsto(u+vj,u+v\ell)$ for $j\neq\ell$ makes
	$I_j=\mathbbm{1}[U\in B(j)]$ pairwise independent Bernoulli$(a)$
	variables. The report density relative to $U$ is at most $e^{\frac{\epsilon}{2}}$
	by Equation~\eqref{eq:pi-ideal-pair-probability}; hence, for $w\in\R^k$,
	\[
		\operatorname{Var}(\langle w,Z\rangle\mid x)
		\leq\mathbb{E}[\langle w,Z-a\mathbf{1}\rangle^2\mid x]
		\leq e^{\frac{\epsilon}{2}}a(1-a)\|w\|_2^2.
	\]
	Dividing by $\Delta^2$ gives the same covariance bound for PI-RAPPOR.
	Both mechanisms also satisfy $\|\mathbf{y}\|_\infty=O\left(\epsilon^{-1}\right)$.

	By norm duality, with $\phi_i(s)=(\langle s,\mathbf{y}_i\rangle)_+$,
	\[
		L=\max_{s\in\{-1,1\}^k}\sum_{i\in C}\phi_i(s).
	\]
	The covariance bound and $\mathbb{E}\mathbf{y}_i=\mathbf{e}_{x_i}$
	give $\mathbb{E}\phi_i(s)=O\left(\frac{\sqrt{k}}{\epsilon}\right)$ uniformly in $s$.
	Symmetrization and contraction for $t\mapsto t_+$~\cite{BartlettM02},
	with independent Rademacher signs $\sigma_i$, give
	\begin{align*}
		\mathbb{E}L
		&\leq O\left(\frac{m\sqrt k}{\epsilon}\right)
		+\mathbb{E}\max_s\sum_{i\in C}(\phi_i(s)-\mathbb{E}\phi_i(s))\\
		&\leq O\left(\frac{m\sqrt k}{\epsilon}\right)
		+O\left(\sum_{j=0}^{k-1}\mathbb{E}
		\left|\sum_{i\in C}\sigma_i y_{i,j}\right|\right)\\
		&=O\left(\frac{m\sqrt k+k\sqrt m}{\epsilon}\right).
	\end{align*}
	The last step uses
	$\mathbb{E}_{\sigma}(\sum_i\sigma_i y_{i,j})^2
	=\sum_i y_{i,j}^2=O\left(\frac{m}{\epsilon^2}\right)$.
	The common lower bounds and total-error argument complete the SD proof.

	For OM, each report has norm $O\left(\frac{k}{\epsilon}\right)$, so replacing $m$ reports
	changes the estimate by $O\left(\frac{mk}{\epsilon}\right)$.
	For a matching lower bound, all corrupted clients submit the all-one
	bit report (the pair $(0,0)$ for PI-RAPPOR). Its decoded vector is
	$b\mathbf{1}$, where $b=\frac{1-a}{\Delta}$ and $b-1=\Theta\left(\epsilon^{-1}\right)$.
	The mean error is $mb\mathbf{1}-\sum_{i\in C}\mathbf{e}_{x_i}$,
	whose norm is $mbk-m=\Omega\left(\frac{mk}{\epsilon}\right)$; Jensen's inequality
	gives the same lower bound on expected error.
	Honest reporting is also an allowed OM strategy and supplies the
	$\frac{k\sqrt n}{\epsilon}$ lower bound.
\end{proof}

\subsubsection{Hadamard Response}
\label{app:hadamard-proof}
\begin{proof}
	Write $\mathbf{h}(y)=H_K[1:k,y]$, so
	$\mathbf{v}_i=\lambda_\epsilon\mathbf{h}(y_i)$.
	Let $A_y$ count corrupted reports equal to $y$, and $A_y^D$ count those
	in $D$. Cauchy--Schwarz and Hadamard orthogonality give, for every $D$,
	\[
		\left\|\sum_{i\in D}\mathbf{h}(y_i)\right\|_1
		\leq\sqrt{k}\|H_K A^D\|_2
		\leq\sqrt{kK}\left(\sum_y A_y^2\right)^{\frac{1}{2}}.
	\]
	Since $\prob{y_i=y}\leq\frac{2}{K}$ and reports are independent,
	\[
		\mathbb{E}\sum_y A_y^2
		=m+\sum_{i\neq i'}\prob{y_i=y_{i'}}
		\leq m+\frac{2m(m-1)}{K}.
	\]
	Jensen's inequality, $K=\Theta\left(k\right)$, and
	$\lambda_\epsilon=\Theta\left(\epsilon^{-1}\right)$ yield
	$\mathbb{E}L=O\left(\frac{m\sqrt{k}+k\sqrt m}{\epsilon}\right)$.
	The common bounds give equality in order and the SD total-error rate.

	Under OM, each decoded column has norm $\lambda_\epsilon k$,
	giving the $O\left(\frac{mk}{\epsilon}\right)$ perturbation bound.
	Submitting column zero for every corrupted client gives mean error
	$m\lambda_\epsilon\mathbf{1}-\sum_{i\in C}\mathbf{e}_{x_i}$,
	of norm $m\lambda_\epsilon k-m=\Omega\left(\frac{mk}{\epsilon}\right)$.
	Jensen's inequality and the strategy in which all clients report honestly give the two
	total-error lower bounds.
\end{proof}

\subsubsection{HST}
\label{app:hst-proof}
\begin{proof}
	A forged bit either equals $y_i$ or $-y_i$, so flipping it changes
	the estimate by $-2\mathbf{v}_i$, twice the change from dropping it.
	Consequently, for every realization,
	\begin{equation}
		\max_{(y_i')_{i\in C}\in\{-1,1\}^m}
		\left\|\lambda_\epsilon\sum_{i\in C}(y_i'-y_i)\mathbf{s}_i\right\|_1
		=2L.
		\label{eq:hst-flip-drop}
	\end{equation}
	Let $S_C$ have columns $\mathbf{s}_i$, $i\in C$.
	Either attack changes the estimate by $\lambda_\epsilon S_Cu$
	with $\|u\|_2\leq2\sqrt m$, so its norm is at most
	$2\lambda_\epsilon\sqrt{km}\|S_C\|_{\mathsf{op}}$.
	Because $C$ is fixed before sampling the public vectors, $S_C$ has
	independent Rademacher entries and
	$\mathbb{E}\|S_C\|_{\mathsf{op}}=O\left(\sqrt{k}+\sqrt m\right)$~\cite{Vershynin18}.
	Thus both perturbations have expected norm
	$O\left(\frac{k\sqrt m+m\sqrt k}{\epsilon}\right)$, proving the total-error upper bounds.
	The common argument proves the SD lower bound.
	Flipping precisely the bits selected by the selective dropout attack
	doubles the sum in Equation~\eqref{eq:sign-total-error}, proving the OM
	attack term. Honest reporting supplies the remaining lower bound.
\end{proof}

\subsection{Proof of Lemma~\ref{lem:rappor-check}}
\label{app:rappor-check-proof}
\begin{proof}
	Outside the input coordinate $x$, the products $y_jz_j$ are independent,
	mean zero, and have interval width at most $b(b-a)$; also
	$|y_xz_x|\leq b^2$. Hoeffding's inequality~\cite{Vershynin18} gives
	\[
		\prob{|\langle\mathbf{y},\mathbf{z}\rangle|\geq b^2+t}
		\leq2\exp\left(-\frac{2t^2}{kb^2(b-a)^2}\right).
	\]
	Set $t=b(b-a)\sqrt{2k\log n}$ to obtain $\frac{2}{n^4}$.
\end{proof}

\subsection{Proof of Theorem~\ref{thm:rappor-consistency}}
\label{app:rappor-consistency}
\label{app:rappor-sampled-check}
\begin{proof}
	The case $n=1$ follows from the deterministic $O\left(\frac{k}{\epsilon}\right)$ error
	bound; assume $n\geq2$.
	Let $\mathcal{H},\mathcal{M}$ index honest and malicious reports.
	By Lemma~\ref{lem:rappor-check}, independence of malicious and honest
	reports, and a union bound, the event $E$ that no pair involving an
	honest report violates the threshold has $\prob{E^c}\leq\frac{2}{n^2}$.
	On $E$, every honest report survives either check.
	If $V$ is the retained set and $B$ counts violating pairs within
	$V\cap\mathcal{M}$, then
	\begin{equation}
		\left\|\sum_{i\in V\cap\mathcal{M}}\mathbf{y}_i\right\|_2^2
		\leq mkb^2+m^2\tau+2kb^2B.
		\label{eq:rappor-retained-norm}
	\end{equation}
	Indeed, diagonal terms are at most $kb^2$, consistent cross terms
	at most $\tau$, and each violating pair contributes at most $2kb^2$.
	The full check has $B=0$.

	For the sampled check, fix reports satisfying $E$ and form the graph
	of violating pairs, all of whose edges join malicious reports.
	A vertex of degree $d_i$ survives only if the uniform sample $S$,
	of size $s=\min\{n,\lceil\frac{n\sqrt{k}}{m}\rceil\}$, misses all its neighbors.
	Thus
	\[
		\prob[S]{i\text{ survives}}
		\leq\frac{\binom{n-d_i}{s}}{\binom ns}
		\leq e^{-\frac{sd_i}{n}},
	\]
	with numerator zero if $s>n-d_i$. Counting surviving edge endpoints,
	\begin{equation}
		\mathbb{E}_S B
		\leq\tfrac12\sum_{i\in\mathcal{M}}d_i e^{-\frac{sd_i}{n}}
		\leq\frac{mn}{2es},
		\label{eq:rappor-surviving-edges}
	\end{equation}
	since $u e^{-\frac{su}{n}}\leq\frac{n}{es}$.
	If $s=n$, then $B=0$; otherwise $\mathbb{E}_S B=O\left(\frac{m^2}{\sqrt{k}}\right)$.

	For either check, Equation~\eqref{eq:rappor-retained-norm},
	$b=O\left(\epsilon^{-1}\right)$, and
	$\tau=O\left(\frac{\sqrt{k\log n}}{\epsilon^2}\right)$ bound the expected squared norm
	of the retained malicious sum, conditional on the fixed reports in $E$,
	by $O\left(\frac{mk+m^2\sqrt{k\log n}}{\epsilon^2}\right)$.
	Cauchy--Schwarz and Jensen's inequality convert this to an $\ell_1$
	bound. Adding the unconditional error from honest clients' reports and the norm (at most $m$)
	of the malicious-input histogram gives
	\[
		\mathbb{E}[\|\widehat F-F(\mathbf{x})\|_1\mathbf{1}_E]
		=O\left(\frac{k\sqrt n}{\epsilon}
		+\frac{mk^{\frac{3}{4}}(\log n)^{\frac{1}{4}}}{\epsilon}\right),
	\]
	using $m\leq n$. On $E^c$, the deterministic error bound
	$nkb+n=O\left(\frac{nk}{\epsilon}\right)$ contributes only $O\left(\frac{k}{n\epsilon}\right)$.
	The sampled check computes at most $ns$ inner products, giving
	$O\left(n^2k\min\{1,\frac{\sqrt{k}}{m}\}\right)$ time.
	The argument uses only $|\mathcal{M}|\leq m$, so also applies when
	the supplied budget is an upper bound.
\end{proof}

\subsection{Proof of Lemma~\ref{lem:had-check}}
\label{app:hadamard-check-proof}
\begin{proof}
	Each count is a sum of independent Bernoulli variables with mean at most
	$\mu=np_1$, where $p_1=\frac{2e^\epsilon}{(e^\epsilon+1)K}$.
	Bernstein's inequality~\cite{Vershynin18} gives
	\[
		\prob{\mathbf{c}(y)_j\geq\mu+t}
		\leq\exp\left(-\frac{t^2}{2\mu+\frac{2t}{3}}\right).
	\]
	For $t=\max\{\mu,5\log(nK)\}$, the exponent is at least
	$\frac{3t}{8}\geq\frac{15}{8}\log(nK)$.
	Since $T_{\mathsf{HR}}=2\mu+5\log(nK)\geq\mu+t$,
	a union bound over $K$ counts gives failure probability at most $\frac{1}{n}$.
\end{proof}

\subsection{Proof of Theorem~\ref{thm:hadamard-consistency}}
\label{app:hadamard-consistency}
\begin{proof}
	Write $D(\mathbf{v})=\lambda_\epsilon H_K[1:k,:]\mathbf{v}$ for the
	decoder, and let $\mathbf{c}_{\mathcal{H}},\mathbf{c}_{\mathcal{M}}$
	be the honest and malicious report histograms.
	With probability at least $1-\frac{1}{n}$, every honest count is at most
	$T_{\mathsf{HR}}$: the preceding Bernstein argument applies to the
	$n-m$ honest reports with the same upper bound $\mu=np_1$.
	On this event $E$, clipping gives a residual
	$\mathbf{r}=\tilde{\mathbf{c}}(y)-\mathbf{c}_{\mathcal{H}}$ with
	$0\leq\mathbf{r}\leq\mathbf{c}_{\mathcal{M}}$,
	$\|\mathbf{r}\|_1\leq m$, and
	$\|\mathbf{r}\|_\infty\leq T_{\mathsf{HR}}$.
	Consequently $\|\mathbf{r}\|_2^2\leq mT_{\mathsf{HR}}$, and orthogonality
	gives
	\[
		\|D(\mathbf{r})\|_1
		\leq\lambda_\epsilon\sqrt{k}\|H_K\mathbf{r}\|_2
		\leq\lambda_\epsilon\sqrt{kKmT_{\mathsf{HR}}}.
	\]
	Adding the unconditional error from honest clients' reports and the malicious-input histogram
	(of norm $m$), using $K=\Theta\left(k\right)$ and
	$T_{\mathsf{HR}}=O\left(\frac{n}{K}+\log(nK)\right)$, yields
	\begin{align*}
		&\mathbb{E}[\|D(\tilde{\mathbf{c}}(y))-F(\mathbf{x})\|_1\mathbf{1}_E]\\
		&\quad=O\left(\frac{k\sqrt n+\sqrt{kmn}
		+k\sqrt{m\log(nK)}}{\epsilon}\right).
	\end{align*}
	On $E^c$, clipping leaves total count at most $n$, so the error
	is $O\left(\frac{nk}{\epsilon}\right)$. Its expected contribution is $O\left(\frac{k}{\epsilon}\right)$,
	absorbed by the error from honest clients' reports.
\end{proof}

\end{document}